\documentclass{aa}
\usepackage[utf8]{inputenc}
\usepackage[varg]{txfonts}
\usepackage{bm}
\usepackage{mathtools}
\usepackage{amssymb}
\usepackage{graphicx}
\usepackage{epstopdf}
\usepackage{natbib}
\usepackage{color}
\usepackage[colorlinks=true,     linkcolor=blue, citecolor=blue, filecolor=blue, urlcolor=blue]{hyperref}

\bibpunct{(}{)}{;}{a}{}{,}

\usepackage{multirow}
\usepackage{booktabs}
\usepackage{stfloats}
\usepackage{placeins}

\newlength{\verticalcompensationlength}
\newcounter{verticalcompensationrows}

\usepackage{lscape}
\usepackage{rotating}
\pdfoutput=1
\begin{document}

   \title{How dust particles orbiting white dwarfs can reveal undetected co-orbiting exoplanets}
   \subtitle{I. Exoplanet on a circular orbit}

 \author{Kyriaki I. Antoniadou\inst{\ref{inst1},\ref{inst2},\ref{inst3}} \and   Dimitri Veras\inst{\ref{inst4},\ref{inst5},\ref{inst6}}
          }

   \institute{Department of Physics, Democritus University of Thrace, 65404, Kavala, Greece
   \\ \email{kantonia@physics.duth.gr}\label{inst1} 
   \and
        Department of Mathematics ``Tullio Levi-Civita'', University of Padua, 35121, Padua, Italy  \label{inst2}
   \and
    Department of Physics, Aristotle University of Thessaloniki, 54124, Thessaloniki, Greece \label{inst3}
   \and
    Centre for Exoplanets and Habitability, University of Warwick, Coventry CV4 7AL, United Kingdom \label{inst4}
   \and
    Centre for Space Domain Awareness, University of Warwick, Coventry CV4 7AL, United Kingdom\label{inst5}
   \and
        Department of Physics, University of Warwick, Coventry CV4 7AL, United Kingdom\label{inst6}
         }

\titlerunning{I. Observational hints about debris}

\authorrunning{Kyriaki I. Antoniadou and Dimitri Veras}

  \abstract
    {Mounting observations of transiting dusty debris orbiting white dwarfs (WDs) display complex signatures that often feature orbital period drifts. The asteroid or planet progenitor of this debris might still be partly intact on a near-circular orbit that is inclined or coplanar, with a period that is similar to that of the dust. Alternatively, the debris could persist concurrently with other intact asteroids or planets that had tidally circularised around the WD at a later time, but still remain directly unobservable.}   
    {We sought to first improve the identification of this detectability threshold by expanding our previous investigation of the WD-secondary-dust co-orbital circular restricted three-body problem (CRTBP) to higher secondary masses corresponding to Ceres, Mercury, Earth, and Neptune. Next, we continued to explore the link between the analytic and numerical characterisations of the CRTBP.}
    {First, we computed families of periodic orbits in the 1/1 mean-motion-resonance (MMR) in both the planar and spatial CRTBP cases, where the dust is on an arbitrarily eccentric orbit. Second, we generated global phase space portraits with dynamical stability (DS) maps. Third, we determined observational links with debris located in very close proximity to WDs via $N$-body simulations.}
   {For each method we applied, we report a novel set of results. These include three newly discovered bifurcation points, $B_T^4, B_{cs}^4$, and $B_{cs}^5$, two branches of asymmetric periodic orbits, called $A_2$ and $A_3$, at high dust eccentricity regime in the 2D-CRTBP, and two families of symmetric orbits in the 3D-CRTBP. The maps reveal stability domains as a function of the dust's eccentricity, argument of pericentre, and slight semi-major axis deviations from the 1/1 MMR. The $N$-body simulations reveal that over 10 yr, mass analogues of Mercury, Earth, and Neptune generate orbital period deviations in the dust of, respectively, $\sim$1-10s, $\sim$10-100~s, and over 100~s.}
   {We identified strong links between analytic periodic orbits, numerical DS maps, and numerical $N$-body simulations. The observed period deviations are easily detectable, aiding in efforts to obtain indirect detections of exoplanets orbiting WDs.}
   \maketitle
\nolinenumbers
%
\section{Introduction} \label{intro}
Periodic orbits represent fundamental structures of $N$-body systems at the location of mean motion resonances (MMRs). These orbits imprint foundational curves in the phase space and also provide direct insights into long-term stability, allowing for credible predictions to be made about the behaviour of celestial bodies in the vicinity of the orbits. 

The exploration of the excitation level and stability of asteroids and planets in the phase space that surrounds these periodic orbits can be carried out using a range of techniques. Across a series of papers, P1 \citep{ave16}, P2 \citep{ave19}, P3 \citep{ave24}, P4 (this work), and P5 \citep{ave26b}, we have used a combination of Lyapunov-based dynamical stability (DS) maps and $N$-body integrations to explore these surrounding regions, while also deriving new families of periodic orbits that guide the phase space exploration. In all these papers, the goal is to investigate white dwarf (WD) planetary systems in the context of the circular or elliptic restricted three-body-problem (CRTBP or ERTBP) with the WD as primary,\footnote{In Stellar dynamics, the nomenclature primary/secondary/third body is preferred to the description of the configuration among two primaries and a third body as major primary/minor primary/secondary body often used in Celestial Mechanics \citep{sze}.} an asteroid or planet as secondary, and a dust particle or an asteroid as the third massless body. Table \ref{tab0} compares the key features of these papers.

Because of their extreme nature, WD planetary systems are particularly well-suited to pushing the boundaries of the existing literature on periodic orbits. Dust, asteroids, and planets can be gravitationally scattered during the WD phase, achieving eccentricities exceeding 0.99 \citep[e.g.][]{brou22}; see \citet{vermusbon2024} for a review. We note that this issue is addressed in both P1 and P2.

After arriving in the close vicinity of the WD due to tidal orbital truncation \citep{verefretal2019,oco20,lietal2025a}, an asteroid or planet will begin to break up. Recent photometric transit observations suggest that this breakup is messy and can occur beyond the WD's Roche radius \citep{vander20,vander21,farihi22,bhaetal2025,guietal2025,heretal2025,Korth2026}. The first transit observations of an asteroid breaking up around a WD \citep{vander15}, WD 1145+017, suggested that the breakup was initially only partial. Here, the parent body co-existed with its resultant ejecta at least for 5-10 years \citep{agd24}. This situation was inferred because of the measurement of transit timing variations, or orbital period deviations, between the ejecta and the parent body. This case is addressed in papers P3, P4, and P5.

In P3, we modelled one specific mass ratio in the CRTBP, with the secondary corresponding to about 10\% of the mass of Ceres, with a mass parameter of $\mu=2\times 10^{-11}$. Although it was well motivated \citep{rapp16,gurr17}, this choice was focussed on systems that specifically resemble WD 1145+017, which does not host any known planets. With JWST, there has been a renewed push to search for planets around white dwarfs \citep{limetal2024,muletal2024,adeetal2025,debetal2025,mauetal2026,muletal2026,pouetal2026} to supplement the census of  six currently known planets \citep{thoarztay1993,sigetal2003,luhburboc2011,ganetal2019,vanetalNat2020,blaetal2021,zhaetal2024}. 

Because of this interest, in this work (P4) and in the companion paper (P5), we have extended the study of P3 to higher secondary masses up to Neptune-mass. Specifically, we utilised four higher mass parameters, normalised with respect to the WD's mass ($0.6 M_\odot$). These normalised mass parameters are $\mu=7.86\times 10^{-10}$ (Ceres), $\mu=2.76\times 10^{-7}$ (Mercury), $\mu=5.005\times 10^{-6}$ (Earth), and $\mu=8.58\times 10^{-5}$ (Neptune). We modelled the dynamics of these three-body problems (TBPs) and, thus, we were able to quantify the extent of the resulting orbital period deviations of the dust.

This paper is organised as follows. We first highlight the fundamental notions of the origin and continuation of periodic orbits at the start of Sect. \ref{POs} before computing the families of periodic orbits in the 1/1 MMR in Sect. \ref{crtbp}. The planet is restricted to a circular orbit, while the dust particle's motion is either coplanar (Sect. \ref{2dcrtbp}) or inclined (Sect. \ref{3dcrtbp}). The novelty with respect to the existing literature is noted in Sect. \ref{novelty}. In Sect. \ref{maps}, we comprehensively characterise the dynamical neighbourhood of such fictitious configurations via DS maps. Then, in Sect. \ref{sims}, we describe the $N$-body simulations we performed and report on the stability and period deviations. We discuss our findings in Sect. \ref{dis} before presenting our conclusions in Sect. \ref{concl}.

\section{The dynamics of 1/1 resonant periodic orbits}\label{POs}
For a brief history of co-orbital dynamics and specific technicalities concerning the computation and characterisation of periodic orbits, we refer to P3 and the references therein. Below, we simply point out only the necessary information with regards to the starting orbits in each type of problem; namely, the bifurcation points, $B$, via
\begin{figure*}[!h]\centering
$\begin{array}{c}
  \includegraphics[width=0.99\textwidth]{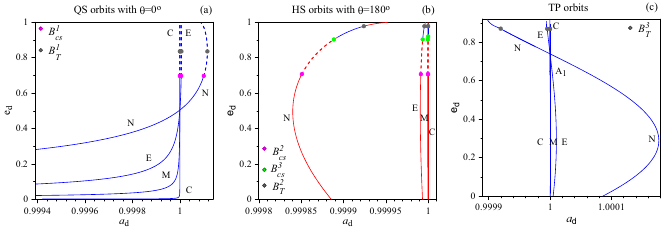}\\
\end{array}$\vspace{-0.3cm}
\caption{Known families of (a) QS, (b) HS and (c) TP (called $A_1$ here) orbits  in the 2D-CRTBP on the $(a_{\rm d},e_{\rm d})$ plane for $a_{\rm a}=1.0$. Stable 1/1 resonant periodic orbits are blue coloured, while the unstable ones are red. Vertical instability is depicted by dashed lines. Magenta and green dots, $B^{1,2,3}_{cs}$, are bifurcation points that generate spatial periodic orbits in the 3D-CRTBP of $xz$-plane and $x$-axis symmetry, called $F$ and $G$, respectively (see Fig. \ref{3DC}). The grey dots, $B^{1,2,3}_T$, are bifurcation points that generate periodic orbits in the 2D-ERTBP (shown in P5). The initials C, M, E, and N stand for the secondary used in each TBP: Ceres, Mercury, Earth, and Neptune.}\vspace{-0.3cm}
        \label{2DC}
\end{figure*}

\vspace{-0.2cm}\begin{itemize}
\item {$B_T$: When the period of the dust particle becomes equal to a multiple of $2\pi$, that is, the period, $T_0$, of the secondary in the CRTBP, we have a bifurcation point that generates orbits in the ERTBP \citep{hach75}. In fact, two branches are generated. Across one branch, the secondary is located at pericentre, while along the other one, the secondary is located at apocentre. Usually, one family is wholly stable and the other is unstable. In P4 (herein), we came across bifurcations generating periodic orbits from and to the following models: from the 2D-CRTBP to the 2D-ERTBP for both symmetric and asymmetric periodic orbits, where we found new families (Sects. \ref{2dcrtbp} herein  and in P5), and from the 3D-CRTBP to the 3D-ERTBP for $xz$-symmetric periodic orbits. Here, we showcase all of these bifurcation points and present the generated families of the 2D-ERTBP and the 3D-ERTBP in P5.}

\item {$B_{s}$: Along the families of planar periodic orbits, we can compute the vertical stability index \citep{hen}, which takes into account the deviations towards the third dimension. When this index gets critical, that is, when we have a transition from vertical stability to instability, the orbit is referred to as "vertical critical" (vco) and a bifurcation point that generates 3D periodic orbits appears. For instance, this scheme was exhibited for various MMRs by \citet{spa}. Here, we encountered a vco, called $B_{cs}$, in the 2D-CRTBP yielding $xz$- and $x$-symmetric periodic orbits in the 3D-CRTBP. In P5, we show the vco of the 2D-ERTBP, called $B_{es}$, generating a family of $xz$-symmetric orbits in the 3D-ERTBP  \citep[see also][for the definition of the $xz$ and $x$-symmetric orbits]{va18}. }

\item {Apart from the vertical stability, we can compute the linear stability of the periodic orbits in the CRTBP \citep{marchal90} and the ERTBP \citep{Broucke1969}. When we have a transition from stability to instability along the families of symmetric periodic orbits, we get bifurcation points that generate asymmetric periodic orbits. This scheme applies to both 2D and 3D TBPs (see e.g. \citealt{voyhadj05,avk11, vta18}). We did not introduce a new symbol for these bifurcation points and we relied instead on the precise visual representation (see Fig. \ref{2DC}).}
\end{itemize}

In Table \ref{tab1}, the bifurcation points arising along in the families of Sect. \ref{POs} are presented. We use the osculating elements to describe the orbits and, in particular, the semi-major axis, $a$, the eccentricity, $e$, the inclination, $i$, the argument of pericentre, $\omega$, mean anomaly, $M$, and the longitude of the ascending node, $\Omega$. We also use the notation $\varpi=\omega + \Omega$ for the longitude of pericentre, $\Delta\varpi$ for the apsidal difference, $\Delta\Omega$ for the nodal difference, and $\lambda=\varpi + M$ for the mean longitude. Therefore, we have adopted the resonant angle $\theta=\lambda_{\rm a}-\lambda_{\rm d}$ \citep{murray} to separate different types of co-orbital motion. In agreement with P3, subscript ${\rm 'a'}$ denotes the secondary body (herein either an asteroid or an exoplanet) and ${\rm 'd'}$ signifies the third body, namely the  dust particle. 

The change reflected in $e_{\rm d}$, as $\mu$ evolved from Ceres to Neptune values, was not significant (i.e. of the order of $10^{-4}$). This dependence becomes significant for greater values of $\mu$  \citep{va18}. In all computations, we set the semi-major axis of the secondary $a_{\rm a}=1$, so that $T_0=2\pi$.

\subsection{Secondary on circular orbit (CRTBP)}\label{crtbp}
We explored all types of orbits in 1/1 MMR: quasi-satellite (QS), horseshoe (HS), and tadpole (TP) orbits (see also P3). The eccentricity of the secondary is exactly equal to 0, $e_{\rm a}=0$ and we let the dust particle move either on the same plane with the secondary (Sect. \ref{2dcrtbp}) or in space on prograde orbits (Sect. \ref{3dcrtbp}). To verify the families of periodic orbits, we used the Bulirsch–Stoer integrator with a tolerance of $10^{-18}$, computed the variation equations and the Poincar\'e surface of section with a tolerance of $10^{-16}$ and the periodic orbits with a tolerance of $10^{-15}$.

\subsubsection{Secondary and dust on coplanar orbits (2D-CRTBP)}\label{2dcrtbp}
In Fig. \ref{2DC}, the known families of planar symmetric (QS and HS) and asymmetric (TP) periodic orbits in the 2D-CRTBP are presented along with the linear horizontal and vertical stability. In panel (a), it becomes evident that the dust particle on a QS orbit cannot move on a circular orbit, as $\mu$ increases from a Ceres-like equivalent asteroid to a Neptune-like exoplanet. A similar result, starting from $\mu\approx 10^{-7}$ and ending at $\mu=0.0477$, where QS orbits become unstable \citep[see][for more details]{HenonGuyot70,Benest74}, was showcased by \citet{Pousse17,Pousse2022}. 

The three grey-coloured bifurcation points, $B_T^1$ in panel (a), $B_T^2$ in panel (b), and $B_T^3$ in panel (c), were found at $T=2\pi$ and generate two branches in the 2D-ERTBP each (included in P5). All of them were shown by \citet{Pousse17} for $\mu=10^{-3}$. The three vcos (magenta and green dots), $B_{cs}^1$, $B_{cs}^2$, and $B_{cs}^3$, were also shown in P3 for $\mu=10^{-11}$. However, here we demonstrate that the semi-major axis of the dust particle varies significantly along them as $\mu$ changes within the regime $\approx [10^{-10}-10^{-5}]$. This information can be used to unravel the phase space (Sect. \ref{maps}) and meticulously explore possible observational hints (Sect. \ref{sims}).  

\begin{figure}
$\begin{array}{l}
  \includegraphics[width=0.99\columnwidth]{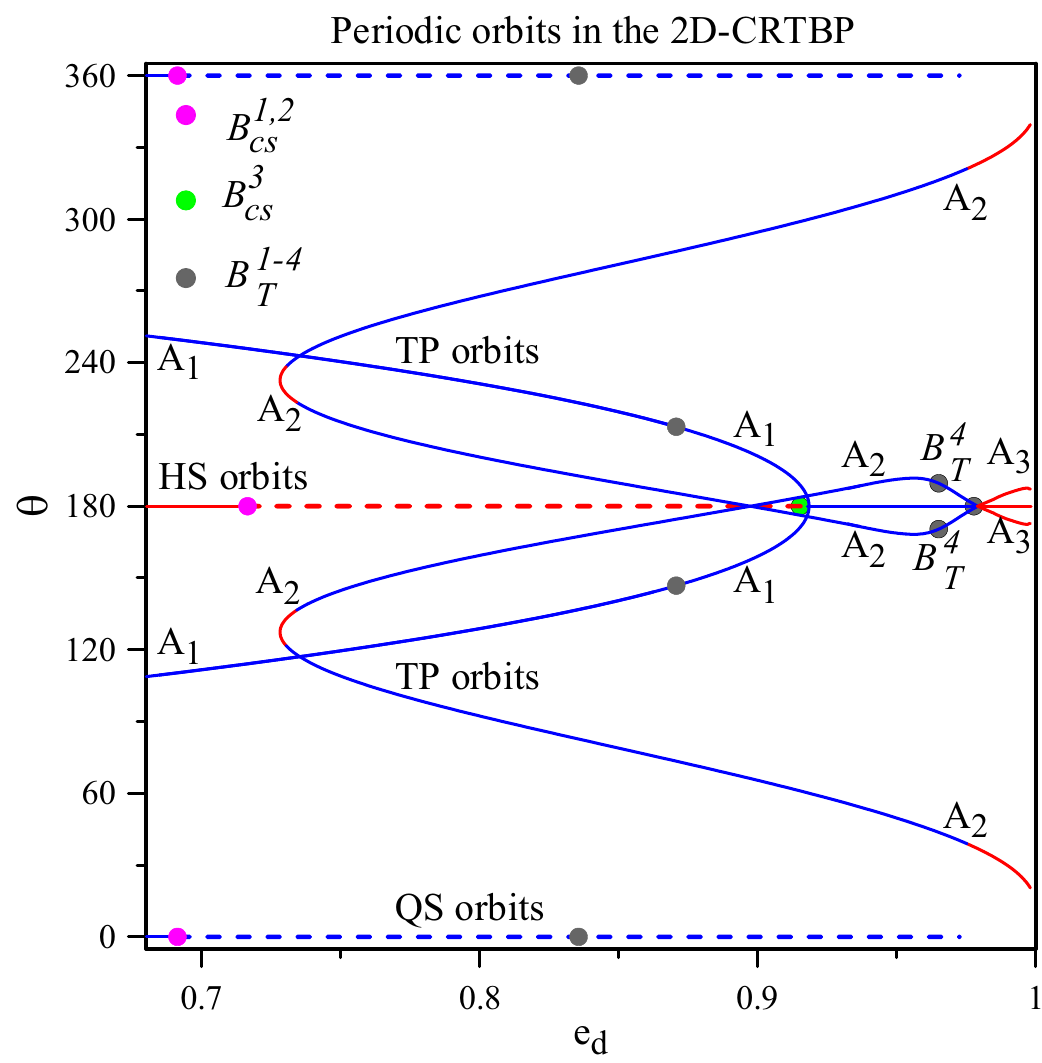}\\
\end{array}$\vspace{-0.3cm}
\caption{New families of highly eccentric asymmetric (TP) periodic orbits, called $A_2$ and $A_3$. $A_2$ (stable) and $A_3$ (unstable) families start after $B_T^2$ and are shown together with the known families of planar symmetric (QS and HS) and asymmetric, $A_1$, (TP) periodic orbits on the $(e_{\rm d},\theta)$ plane in the 2D-CRTBP. The new bifurcation point along $A_2$ family, $B_T^4$, generates new families of asymmetric periodic orbits in the 2D-ERTBP (see P5). Presentation is the same as in Fig. \ref{2DC}.}\vspace{-0.3cm}. 
        \label{2DC_all}
\end{figure}

\begin{figure}
$\begin{array}{c}
  \includegraphics[width=0.99\columnwidth]{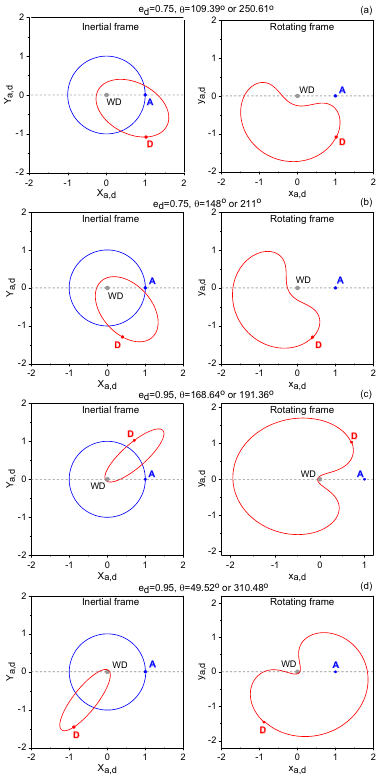}\\
\end{array}$\vspace{-0.3cm}
\caption{Selection of stable asymmetric periodic orbits along the new $A_2$  family in the 2D-CRTBP shown in Fig. \ref{2DC_all}, integrated for one period, $T=2\pi$, and presented in the inertial (\textit{left}) and the rotating (\textit{right}) frame of reference. The initial conditions are reported above each row, while $A$ and $D$ denote the secondary and the dust particle.}\vspace{-0.3cm}
        \label{iraorbits}
\end{figure}

\begin{figure*}
$\begin{array}{c}
  \includegraphics[width=0.99\textwidth]{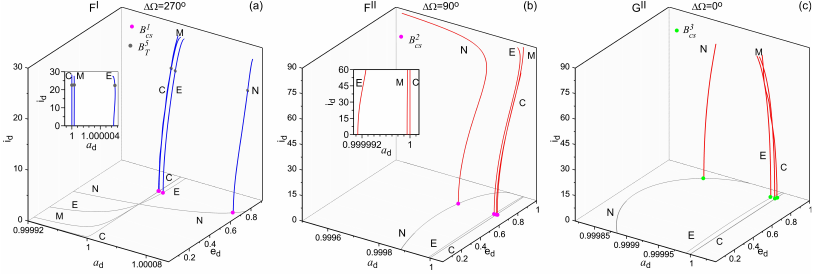}\\
\end{array}$ \vspace{-0.3cm}
\caption{Families of 3D symmetric QS and HS periodic orbits in the $(a_{\rm d},e_{\rm d},e_{\rm d})$ space in the 3D-CRTBP with $a_{\rm a}=1$ and $i_{\rm a}=0^{\circ}$. The QS orbits of the families $F^I$ in (a) have a nodal difference $\Delta\Omega=270^{\circ}$, while the HS orbits, $F^{II}$ and $G^{II}$ in (b) and (c) have a nodal difference $\Delta\Omega=90^{\circ}$ and $\Delta\Omega=0^{\circ}$, respectively. The bifurcation point $B_T^5$ (grey dot) generates two families of periodic orbits in the 3D-ERTBP discussed in P5. The families of the 2D-CRTBP, reported in Fig. \ref{2DC}, are shown in grey colour on the ($a_{\rm d},e_{\rm d}$) plane. Otherwise, presentation is the same as in Fig. \ref{2DC}.}\vspace{-0.3cm}
        \label{3DC}
\end{figure*}

In Fig. \ref{2DC_all}, we provide a global view of all families in the 2D-CRTBP, as shown in P3, while focussing on the region where the two new families of asymmetric periodic orbits exist. The known family $A_1$ starts ($e_{\rm d}=0$) at $L_4$ (or $L_5$) and ends at an HS orbit, where the linear horizontal stability changes. We note that the $B_{cs}^3$ vco (green dot on an HS orbit; see also Fig. \ref{2DC}b) exists at a lower $e_d$ value. 

The new families of asymmetric (TP) orbits, called $A_2$ and $A_3$, appear at very high eccentricity values of the dust particle, start from an HS (symmetric) orbit after the bifurcation point $B_T^2$ (grey dot at $e_{\rm d}=0.977$), where the linear stability changes at $e_{\rm d}=0.982$. Both terminate at a collision with the WD ($e_d\rightarrow 1$). More precisely, the $A_2$ family starts with stable (blue coloured) orbits and becomes unstable (red) before its termination close to QS orbits ($\theta\approx 20^{\circ} {\rm or\,} 340^{\circ}$), while $A_3$ exhibits an almost discernible segment of unstable orbits and ends close to HS orbits ($\theta\approx 180^{\circ}$). In Fig. \ref{iraorbits}, we showcase stable asymmetric periodic orbits along the new $A_2$  family. Furthermore, the $A_2$ family possesses a new bifurcation point, $B^4_T$, at $e_{\rm d}=0.964$ which generates new families of asymmetric orbits in the 2D-ERTBP (shown in P5). The discovery of those new families in the ERTBP, at first hand, incited the continuation in a backward direction and enabled us to identify the $A_2$ and $A_3$ families.

\subsubsection{Secondary and dust on inclined orbits (3D-CRTBP)}\label{3dcrtbp}
In Fig. \ref{3DC}, the families of spatial symmetric (QS and HS) periodic orbits in the 3D-CRTBP are presented for the mass parameter regime $[10^{-10}-10^{-5}]$. Similarly to the 2D-CRTBP, the semi-major axis of the dust particle varies significantly as $\mu$ changes, and we provide magnifications, whenever this is not perceptible. 

In panel (a), the orbits of the $F^I$ families extend up to $i_{\rm d}=30^{\circ}$, while in panels (b) and (c), the families $F^{II}$ and $G^{II}$ could have retrograde orbits, which are beyond the scope of this study. Only one bifurcation point is found along the three groups of families. Then, $B^5_T$ (grey dot in panel a) is a starting (or terminating) orbit, where two branches of the 3D-ERTBP would join together, as demonstrated in P5 and \citet{va18}.

For reasons of completeness, we note that two new families in the 3D-CRTBP were also found. Both of them bifurcate from the vco, $B_{cs}^4$ and $B_{cs}^5$, that exist along the circular family \citep[see][for more details]{spis} close to the 1/1 MMR. Both of them consist of unstable, low-eccentricity $(e_d < 0.12)$ periodic orbits and exist for low inclination values $(i_d < 12^{\circ})$. Therefore, we chose to present them in Fig. \ref{circfam}, since such a configuration would unlikely accumulate dust for observable timescales.

\subsection{Novelty with respect to the existing literature on families of periodic orbits in the co-orbital dynamics}\label{novelty}

With regard to the 2D dynamics, all known families of QS, HS, and TP orbits are called family $f$, $b$ or $L_3$, and $\mathcal{L}_4^s$ (or $\mathcal{L}_5^s$) in the existing literature \citep[see e.g.][]{jackson1913,deprit67,Broucke68,Henon69,HenonGuyot70,Benest74,Bruno94,hen97,Pousse17}, and \citet{Pousse2022}. These families are shown here for the  $\mu$ values in the regime $[10^{-10}-10^{-5}]$. Along with the linear stability, we also computed the vertical stability and showcase the vertical critical orbits. Concerning the 3D dynamics, all known families in Fig, \ref{3DC}, were computed for  $\mu$ values in the regime $[10^{-10}-10^{-5}]$ until their termination. 

The novelty in the 2D-CRTBP is the discovery of two families of asymmetric (TP) periodic orbits, called $A_2$ and $A_3$ (Fig. \ref{2DC_all}), which start at $e_{\rm d}=0.982$ (after $B_T^2$) on the family of HS orbits, where a stability transition is admitted (red segment). Both families terminate at a collision with the WD ($e_d\rightarrow 1$). We call these new branches the $A_2$ and $A_3$ families. These families are not linked with the short- nor the long-period tadpole orbits, called $\mathcal{L}_4^l$ (or $\mathcal{L}_5^l$) in the existing literature, since the latter start from $L_4$ (or $L_5$) and terminate on the family of short-period tadpole orbits, $\mathcal{L}_4^s$ (or $\mathcal{L}_5^s$); this family is called $A_1$ herein in Figs. \ref{2DC}c and \ref{2DC_all} \citep[see e.g.][for details on the morphology of the full families (mono-parametric curves) of the short- and long-period TP orbits]{Goodrich,deprit67}. The families described as Type I and Type II by \citet{Goodrich} belong to the same family of short-period TP orbits ($A_1$ family), but differ in the type of section the Poincar\'e map is computed for. Furthermore, the new $A_2$ family of TP orbits possesses a new bifurcation point, $B_T^4$, which generates new families in the 2D-ERTBP, as shown in P5.

As for the 3D-CRTBP, the novelty reflects on two new families of 3D symmetric periodic orbits shown in Fig. \ref{circfam}a, which bifurcate from two vcos of the circular family, $B^4_{cs}$ and $B^5_{cs}$ (Fig. \ref{circfam}b,c) . These new families consist of $x$-symmetric and $xz$-symmetric periodic orbits and both are wholly unstable.

\section{Phase space visualisation}\label{maps}
We computed DS maps with the use of the detrended fast Lyapunov indicator (DFLI), which has long been established as reliable and accurate in celestial mechanics \citep{voyatzis08}. The maximum integration time, $t<t_{\rm max}$, for each point in the grids corresponded to almost five billion orbits of the dust particle. This integration time was deemed sufficient for $\mu$ values below $10^{-3}$ (Jupiter). We used the Bulirsch–Stoer integrator with a tolerance of $10^{-14}$ and allowed the computations to proceed while DFLI$(t)$ was under either 30 or $t_{\rm max}$. Results with ${\rm DFLI}<2$ signalled stable orbits, while outputs with ${\rm DFLI}>15$ indicated chaoticity, as the DFLI continues to increase steeply thereafter \citep{numan2014}. 

\begin{figure*}[!h]\centering
$\begin{array}{c}
\includegraphics[width=0.99\textwidth]{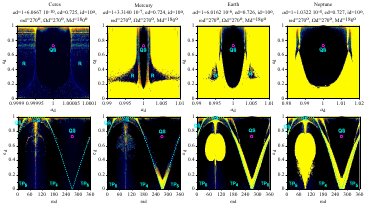}\\\includegraphics[width=0.99\textwidth]{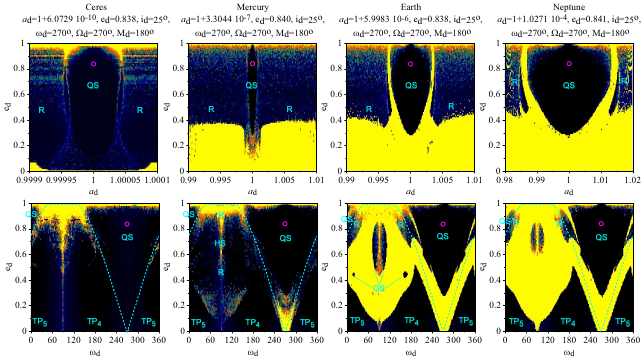}\vspace{-0.2cm}\\
\hspace{0.3cm}\includegraphics[width=0.18\textwidth]{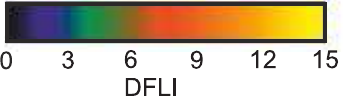}\end{array}$
\caption{Pairs of DS maps for $\Delta i=10^{\circ}$  (\textit{top}) and $\Delta i=25^{\circ}$ (\textit{bottom}) guided by a stable periodic orbit (magenta-coloured circle) of the $F^{I}$ family. Cyan dashed lines correspond to planetary collisions and close encounters. The chosen initial conditions are reported above each column. The resonant angle can be computed via $\omega_{\rm d}$, as $\theta=-\omega_{\rm d}-90$. The libration of $\theta$ around $0^{\circ}$ and $180^{\circ}$ is marked with QS and HS, respectively, while asymmetric libration is denoted by TP and rotation by R.}\vspace{-0.3cm}
\label{mFI}
\end{figure*}

We present the captures of different planes of the phase space in the neighbourhood of specific 3D periodic orbits shown in Fig. \ref{3DC}. For each secondary (each $\mu$ value) and family ($F^I$, $F^{II}$ and $G^{II}$), we chose  mutual inclination values ($\Delta i=10^{\circ}, 25^{\circ}, 45^{\circ}\,{\rm and}\, 85^{\circ}$). We initiated the DS maps (Fig. \ref{mFI}, and Figs. \ref{mFII}-\ref{2DC_S10}). For each DS map, only a couple of orbital elements vary either on the $(a_{\rm d}, e_{\rm d})$ or on the $(\omega_{\rm d}, e_{\rm d})$ plane, while the rest remain fixed. In all cases, $a_{\rm a}=1.0$, $e_{\rm a}=0.0$, $i_{\rm a}=0.0^{\circ}$, $\omega_{\rm a}=M_{\rm a}=\Omega_{\rm a}=0.0^{\circ}$. Therefore, in the DS maps, the resonant angle is $\theta=-\omega_{\rm d}-\Omega_{\rm d}-M_{\rm d}$ and the mutual inclination between the secondary and the dust particle is  $\Delta i$=$i_d$. The rest orbital elements of the  periodic orbits are reported above each pair of planes per secondary.  

The dark regions correspond to long-term stable orbits, the pale ones to chaotic orbits, and the white domains to orbits where the numerical integration failed at $t<t_{\rm max}$ due to very close encounters. Each time, we monitored the resonant angle, $\theta$, and denoted its libration about $0^{\circ}$ with QS, $180^{\circ}$ with HS and any other value with TP followed by a subscript of 4 or 5 denoting asymmetric libration mandated by $L_4$ or $L_5$. Areas where $\theta$ rotated are indicated by R. Examples of these classifications can be found in Fig. \ref{thetaa}, where we showcase the equilibria (periodic orbits) and the regular or irregular motion around them (i.e. dark-coloured domains) on the $(\theta, a_{\rm d})$ plane. Orbits enclosing both TP and QS domains are marked with `QS, TP' in the bottom row of Fig. \ref{mFII} (also see the top-middle panel of Fig. \ref{thetaa}, and \citealt{Namouni99} and \citealt{Nesvorny02}). We also delineated the planetary collisions and close encounters with dashed cyan curves. Finally, the dynamical separatrices dividing motion into regions where $\theta$ is librating and rotating and circulating are computationally colour-coded via the DS maps.
 
\section{$N$-body simulations}\label{sims}

We went on to perform $N$-body simulations by using the IAS15 integrator in the REBOUND simulation package \citep{rein2012,rein2015}. As in P3, here we report the maximum variation in the orbital period of the dust particle, which is directly observed, as well as collisions with the secondary.

Because our aim is to make a direct comparison with P3, we adopted the same value of $a_{\rm a}$=0.012~{\rm au} in this work. That value corresponds to a location with an orbital period of around 15 hours, motivated by the initial observations of periodicities seen in the WD~1145+017 system \citep{vander15,rapp16}.

We then needed to rescale this value of the semi-major axis to what was used to compute the periodic orbits. The necessary transformation was introduced in P1 (also featured in P2 and P3) and it is achieved by multiplying by the factor $\sqrt[3]{0.600000000786}$ for Ceres, $\sqrt[3]{0.600000276}$ for Mercury, $\sqrt[3]{0.600005005}$ for Earth, and $\sqrt[3]{0.6000858}$ for Neptune. 

The radii of the planets also require a similar transformation. We fed into this transformation the actual radii of Mercury, Earth, and Neptune, as well as a radius for our Ceres-like secondary that corresponds to a spherical object with Ceres' mass and a density of $2$~g/cm$^3$. The importance of computing these radii lies with collision detection with our test particles, which are represented by dust particles.

We also implemented collision detection with the WD. We treated this collision as any event where a test particle encountered the WD Roche sphere, whose spatial extent is approximated by $1 R_\odot$. That choice corresponds to P3; in turn, we limited the highest $e_{\rm d}$ value we could sample to about 0.6.

In our non-coplanar cases, we sampled mutual inclinations of 10, 25, 45, and 85 degrees. We chose these values because they all exceeded a few degrees, which is the maximum critical mutual inclination beyond which a planet may remain hidden from our line of sight with transit observations.

We set the minimum timestep of the IAS15 integrator (i.e. an adaptive timestep scheme) to 63.1 sec. That value corresponds to $1/40^{th}$ of an orbit with a period of 42.1 minutes or a semi-major axis of 0.0016 au, which is well within the WD's Roche sphere. We ran our simulations for 10 yr, corresponding to roughly the time since the acquisition of the first transit observations of debris orbiting WDs \citep{vander15} and at least twice the time that has elapsed since similar observations of subsequently discovered systems \citep{vander20,vander21,farihi22,bhaetal2025,guietal2025,heretal2025,Korth2026}.

Our $N$-body plots provide statistical averages based on many values of $M$. We emphasise the difference with our DS maps, which were computed by using single initial values for the $M$. Additionally, the timescales over which both types of results are presented differ by several orders of magnitude; namely, 10 yrs and 2.5 Myrs integration time for the simulations and the DS maps, respectively.

With regard to the grids on the $(a_{\rm d} - a_{\rm a},e_{\rm d})$ and the $(\omega_{\rm d},e_{\rm d})$ planes, each component was sampled uniformly. The initial conditions were derived from the families of periodic orbits computed in Sect. \ref{POs}, and these coincide with the ones chosen for the computation of the DS maps in Sect. \ref{maps} (reported in Figs. \ref{mFI}-\ref{mGII}). We present our $N$-body simulation results in Figs. \ref{sFI_we}-\ref{cMENweae} and \ref{sFII_ael}-\ref{sGII_wer}. 

\subsection{Period deviations}\label{perdev}

\begin{figure*}[t]
\centering
\sidecaption
$\begin{array}{cc}
\includegraphics[width=6cm]{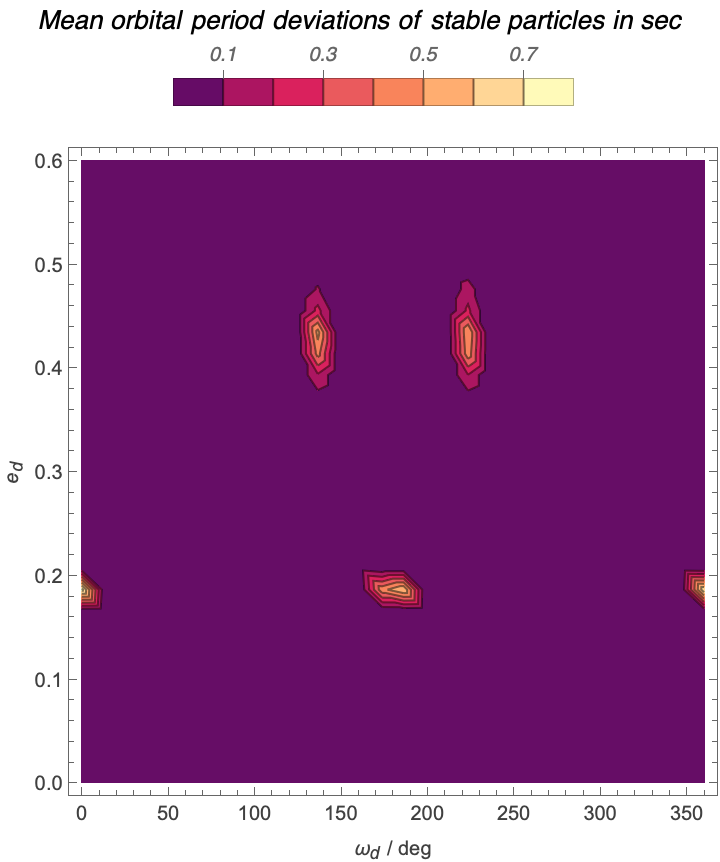}&\includegraphics[width=6cm]{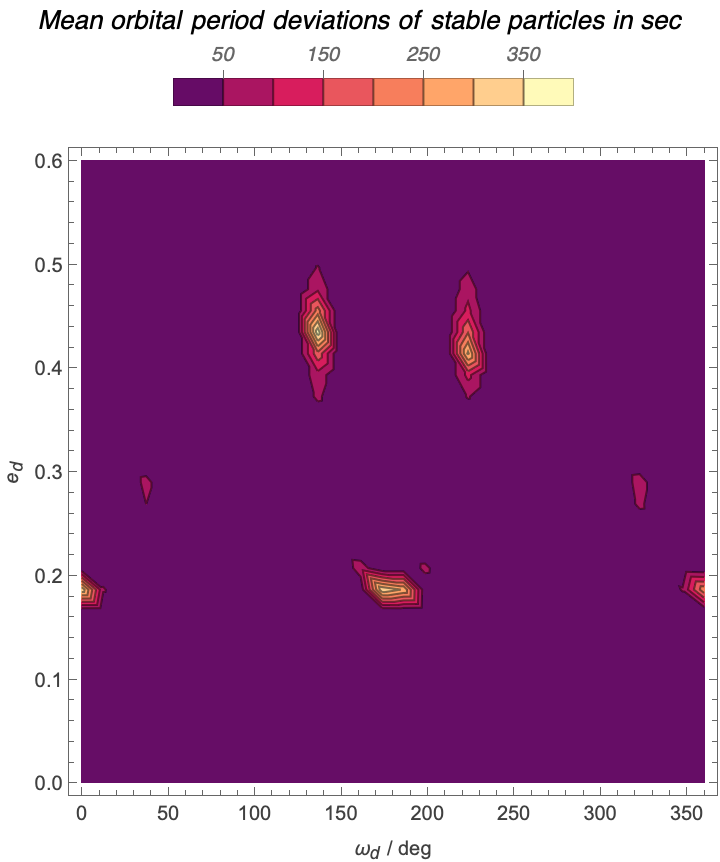}\\\includegraphics[width=6cm]{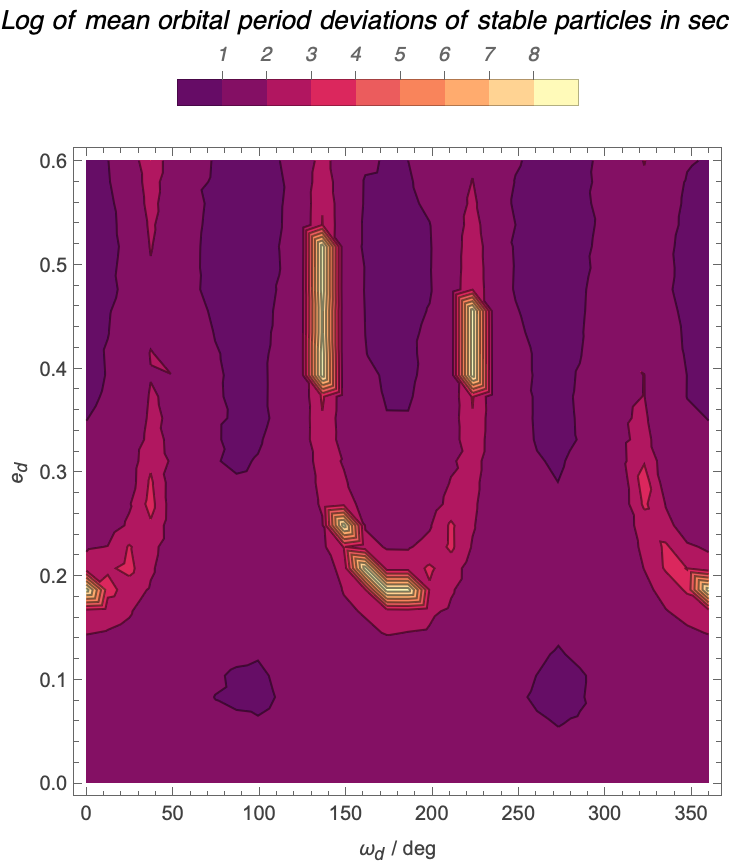}&\includegraphics[width=6cm]{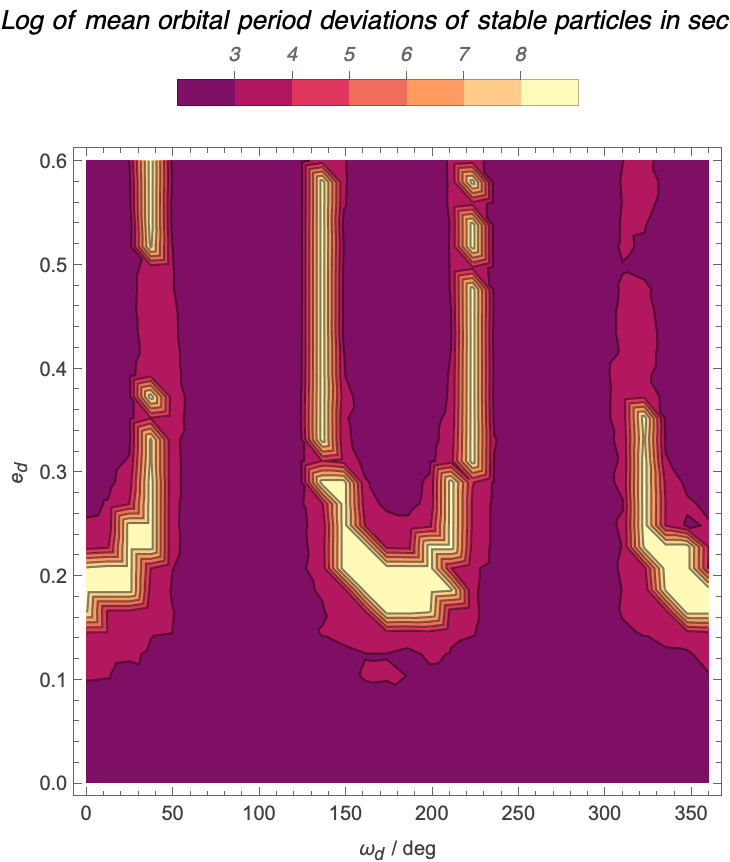}
\end{array}$
\caption{Simulations over a 10 yr timespan. The initial conditions were extracted by a stable periodic orbit of the $F^{I}$ family with $\Delta\Omega=270^{\circ}$ and $\Delta i=25^{\circ}$. The orbital period deviations are colour-coded on the $(\omega_{\rm d},e_{\rm d})$ plane for Ceres (\textit{top-left}), Mercury (\textit{top-right}), Earth (\textit{bottom-left}), and Neptune (\textit{bottom-right}).}
\label{sFI_we}
\end{figure*}

\begin{figure*}[t]
\centering
\sidecaption
$\begin{array}{cc}
\includegraphics[width=6cm]{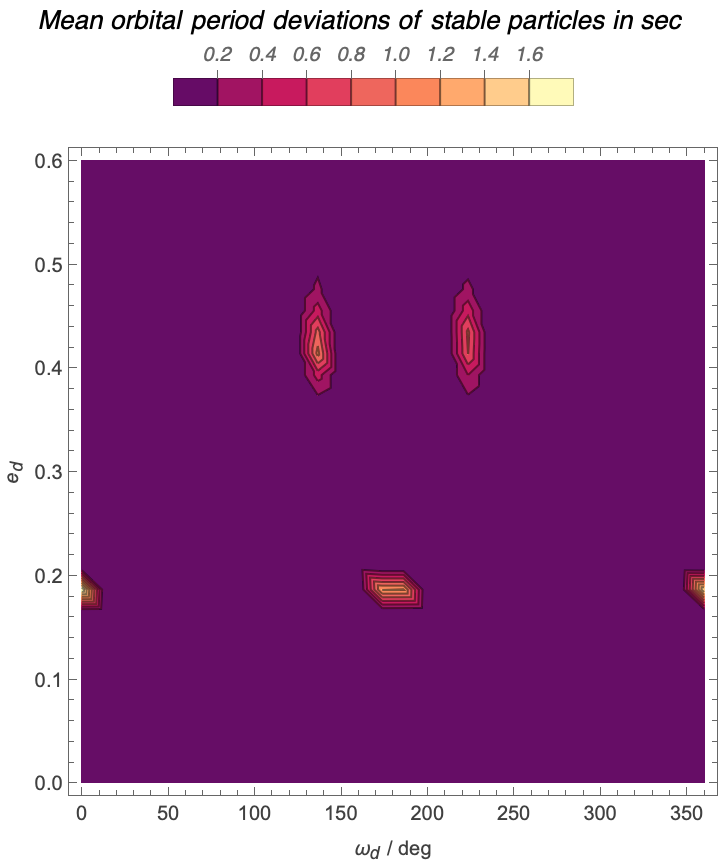}& \includegraphics[width=6cm]{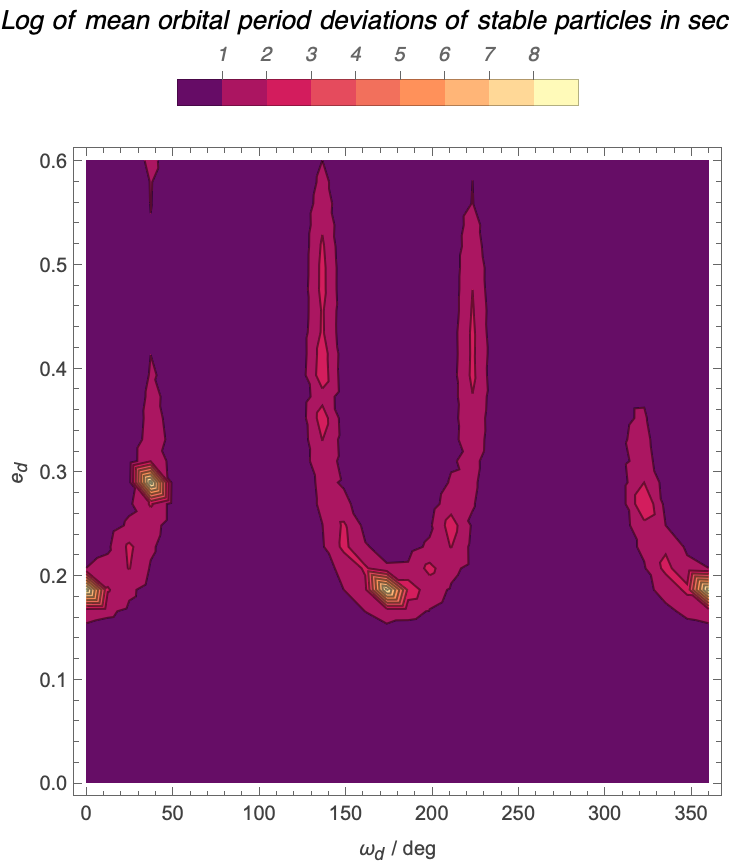}\\\includegraphics[width=6cm]{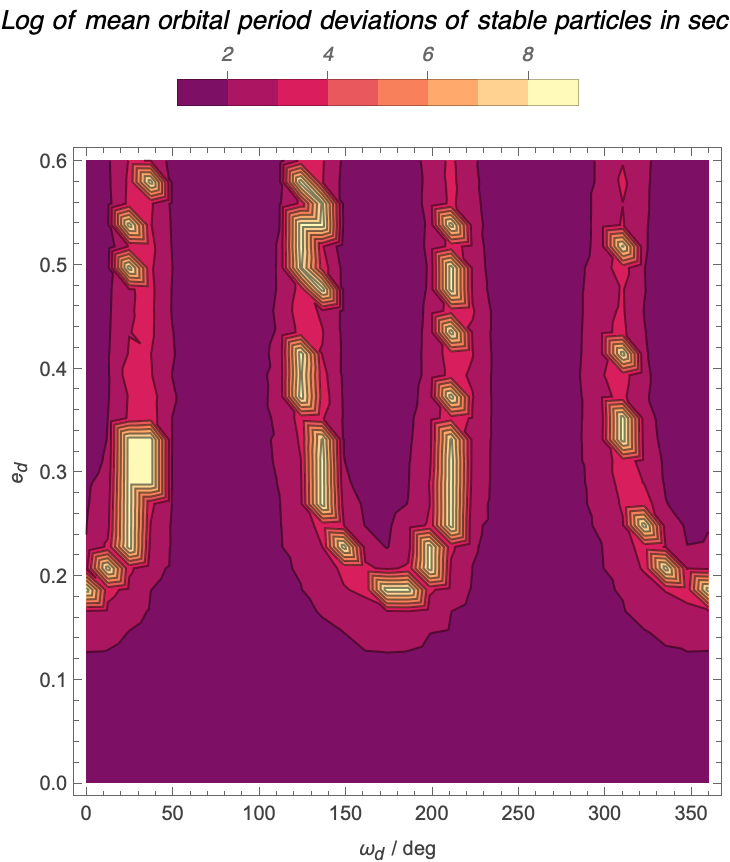}& \includegraphics[width=6cm]{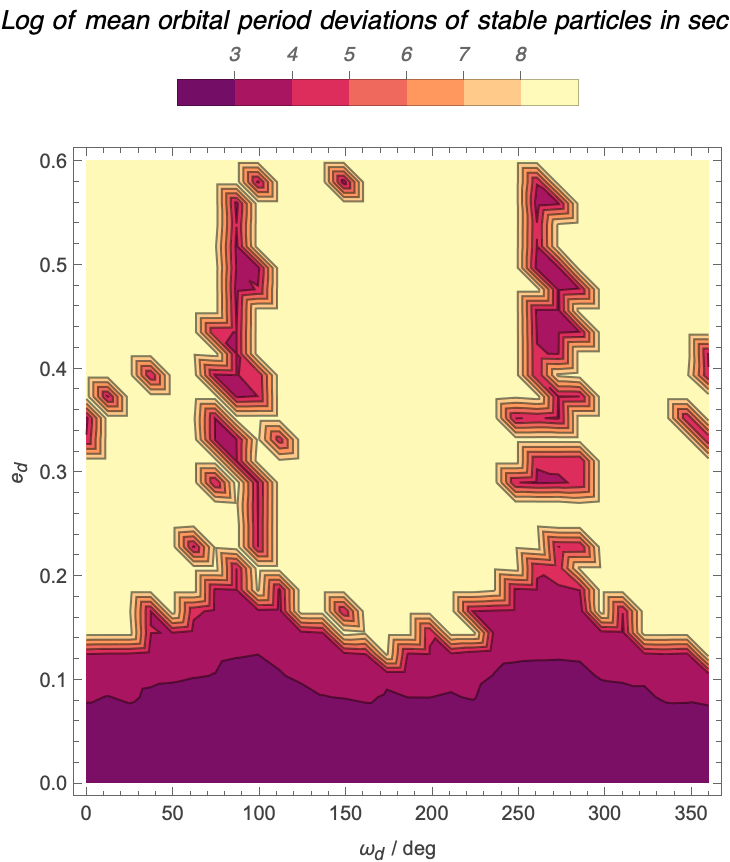}
\end{array}$
\caption{Simulations with initial conditions extracted by an unstable periodic orbit of the $F^{II}$ family with $\Delta\Omega=90^{\circ}$ and $\Delta i=10^{\circ}$. Presentation is the same as in Fig. \ref{sFI_we}.}
\label{sFII_wel}
\end{figure*}

\begin{figure*}[t]
\centering
\sidecaption
$\begin{array}{cc}
\includegraphics[width=6cm]{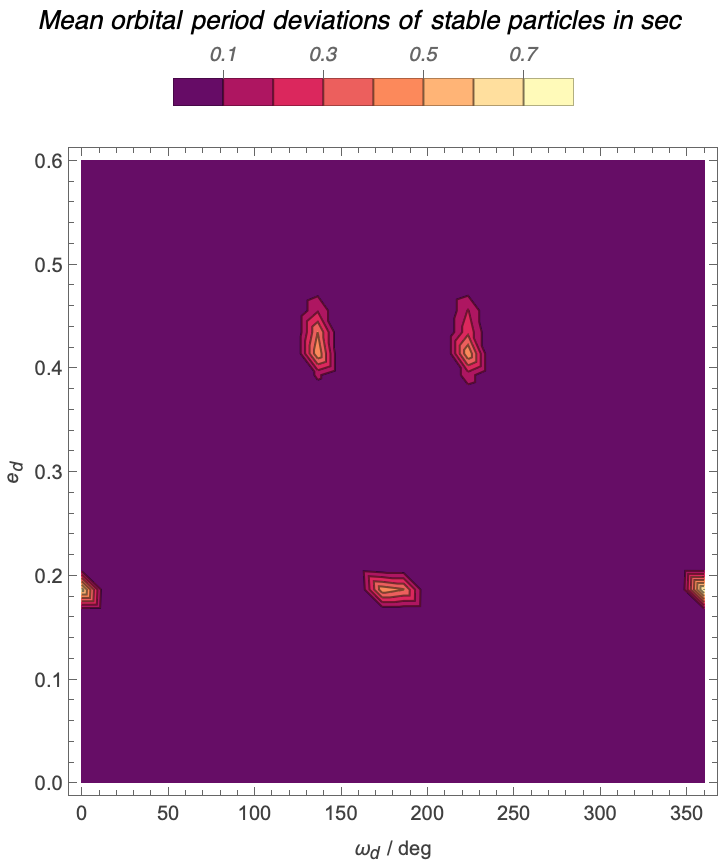}& \includegraphics[width=6cm]{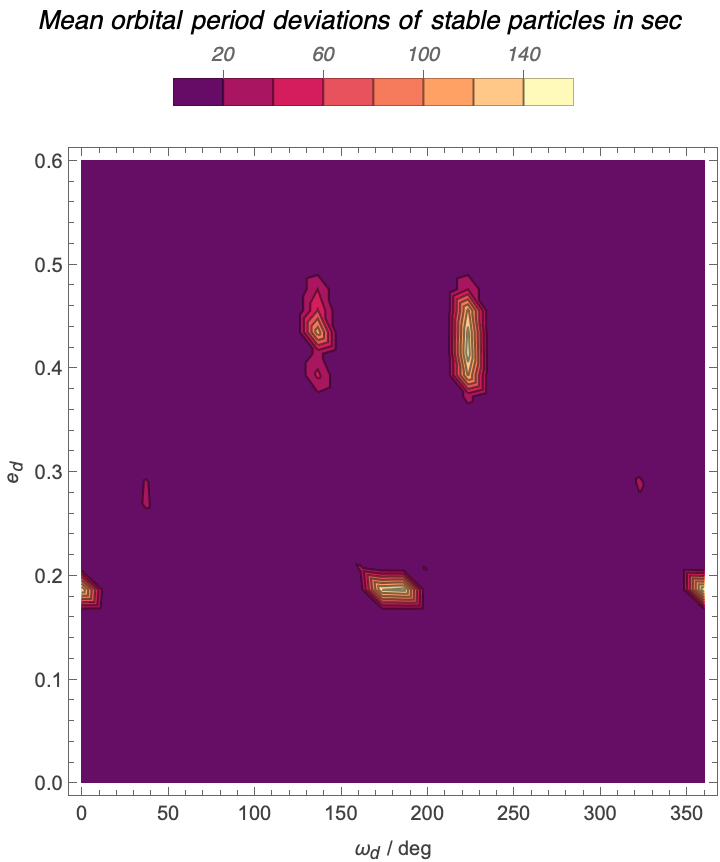}\\\includegraphics[width=6cm]{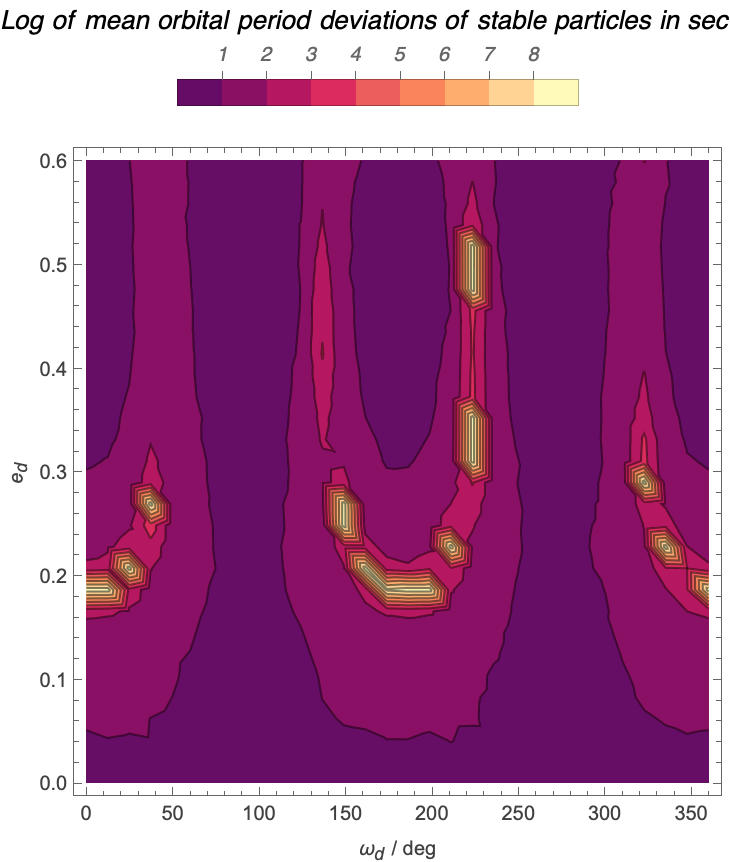}& \includegraphics[width=6cm]{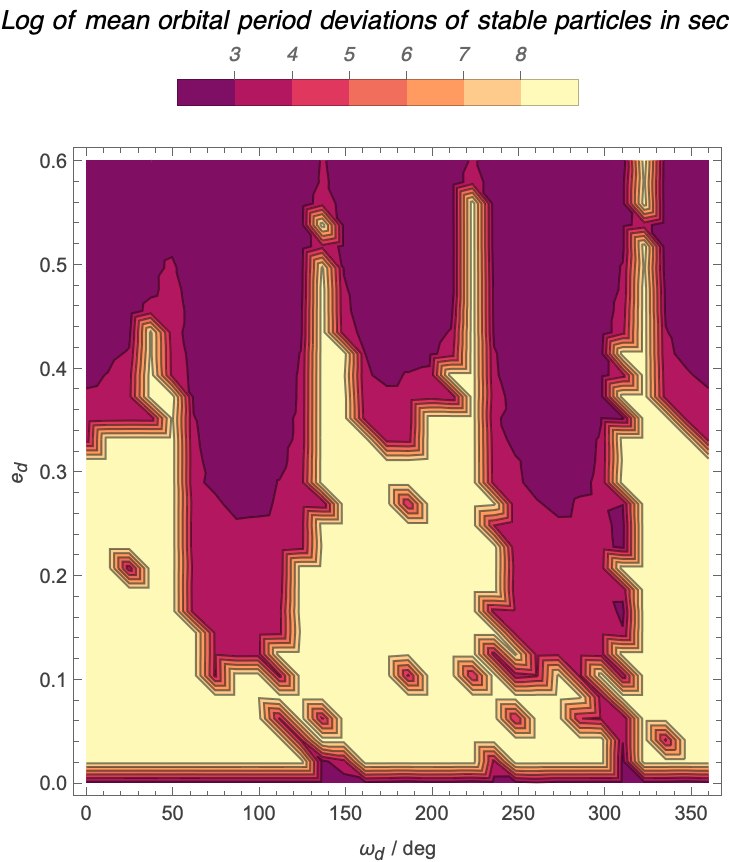}
\end{array}$
\caption{Simulations with initial conditions extracted by an unstable periodic orbit of the $F^{II}$ family with $\Delta\Omega=90^{\circ}$ and $\Delta i=45^{\circ}$. Presentation is the same as in Fig. \ref{sFI_we}.}
\label{sFII_wer}
\end{figure*}

\begin{figure*}[t]
\centering
\sidecaption
$\begin{array}{cc}
\includegraphics[width=6cm]{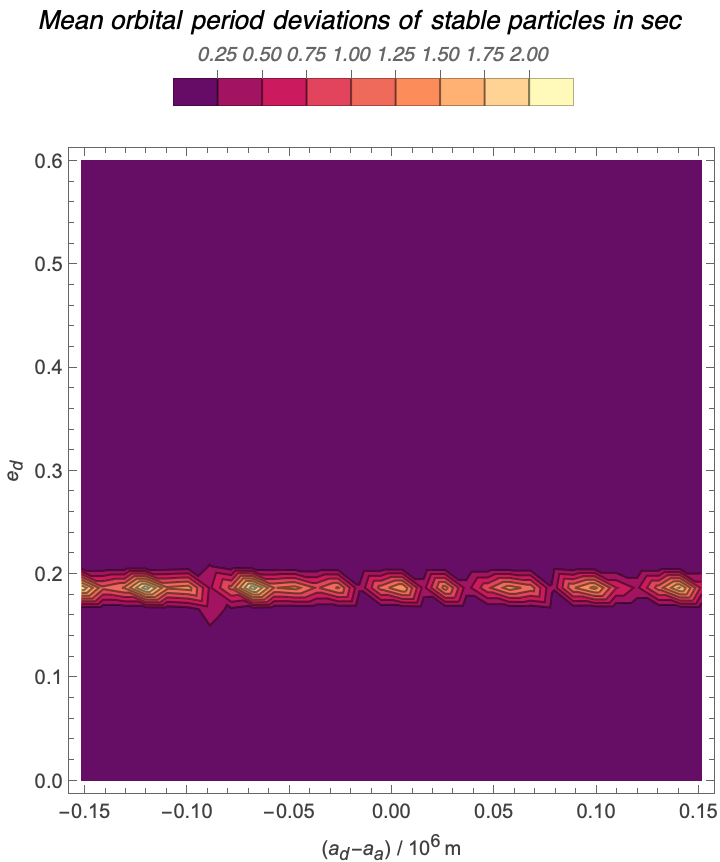}& \includegraphics[width=6cm]{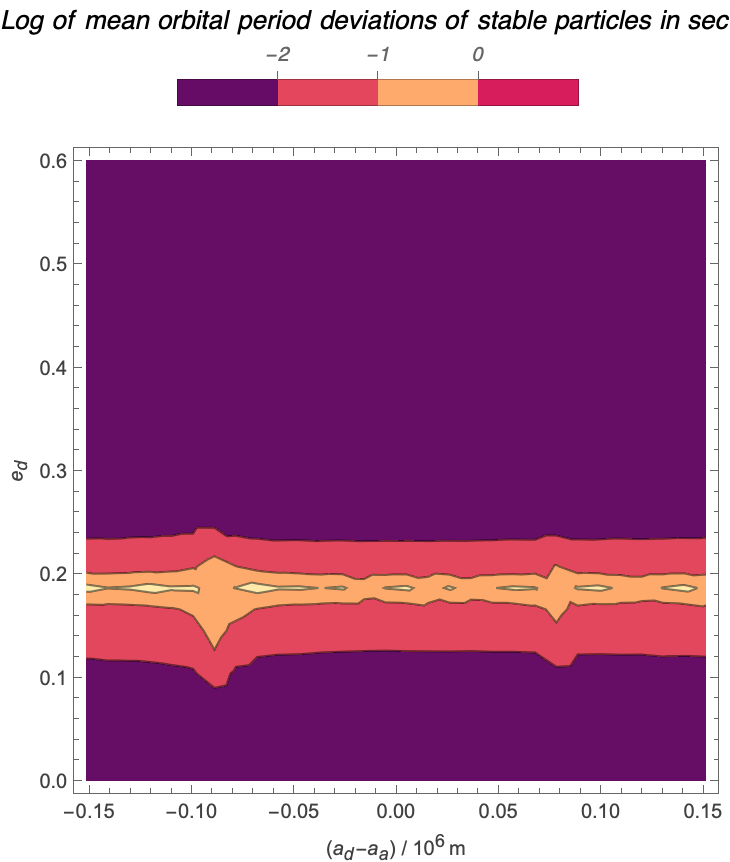}\\\includegraphics[width=6cm]{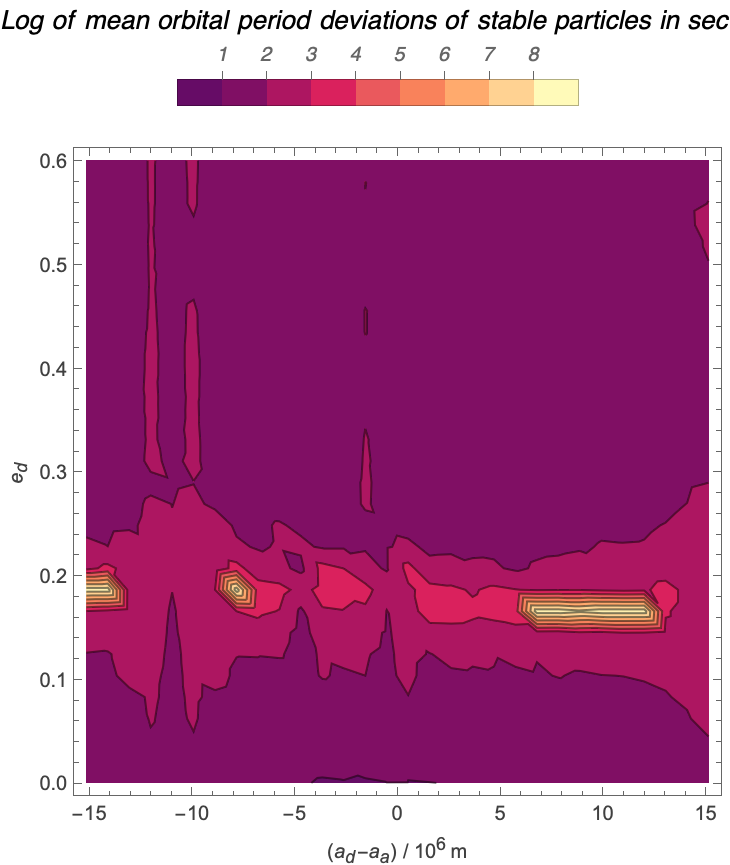}& \includegraphics[width=6cm]{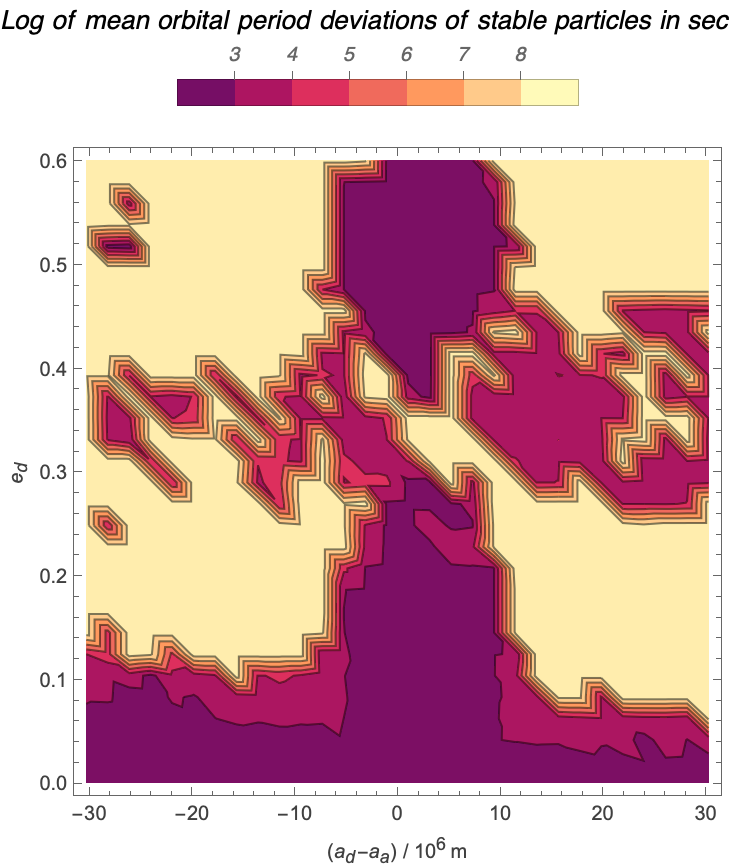}
\end{array}$
\caption{Simulations with initial conditions extracted by an unstable periodic orbit of the $G^{II}$ family with  $\Delta\Omega=0^{\circ}$ and $\Delta i=10^{\circ}$ on the $(a_{\rm d} - a_{\rm a},e_{\rm d})$ plane. Presentation is the same as in Fig. \ref{sFI_we}.}
\label{sGII_ael}
\end{figure*}

\begin{figure*}[t]
\centering
\sidecaption
$\begin{array}{cc}
\includegraphics[width=6cm]{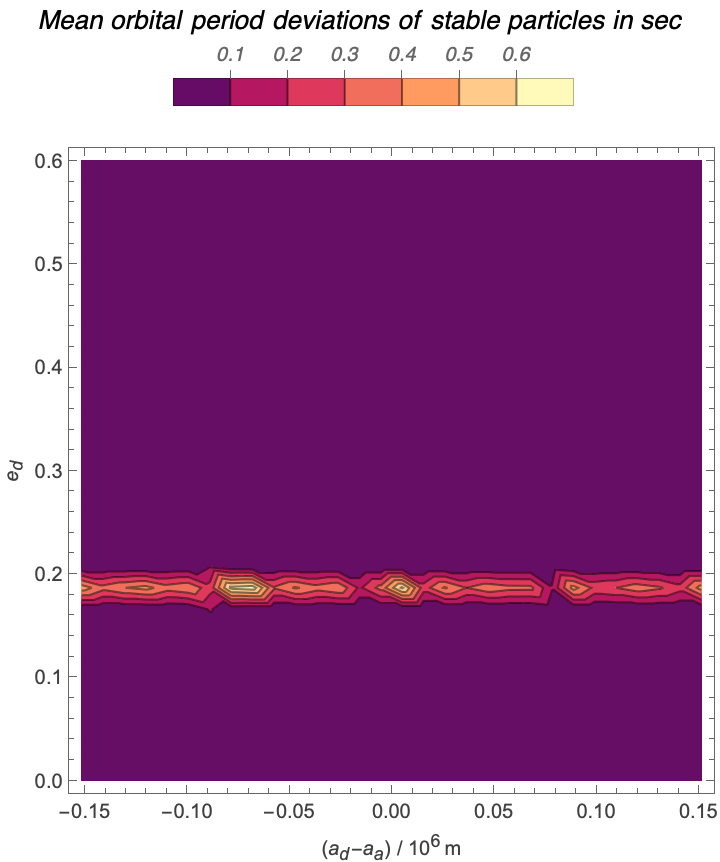}& \includegraphics[width=6cm]{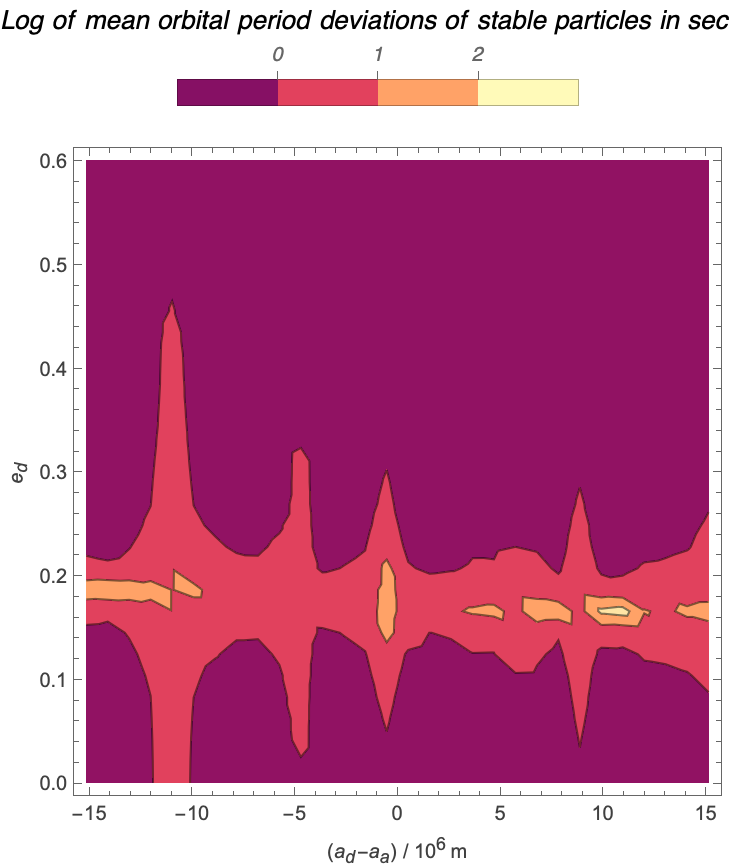}\\\includegraphics[width=6cm]{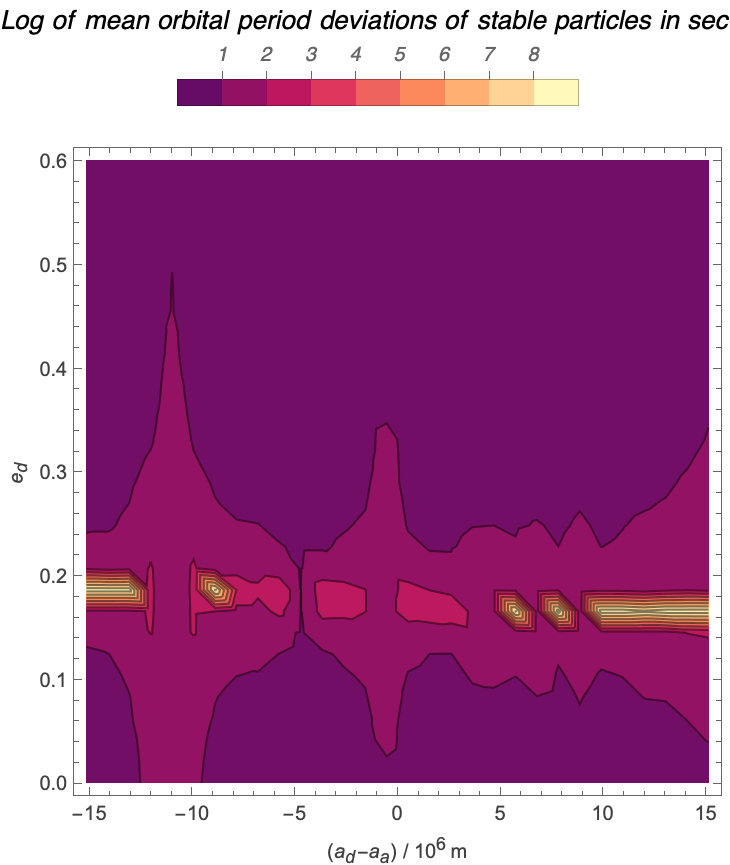}& \includegraphics[width=6cm]{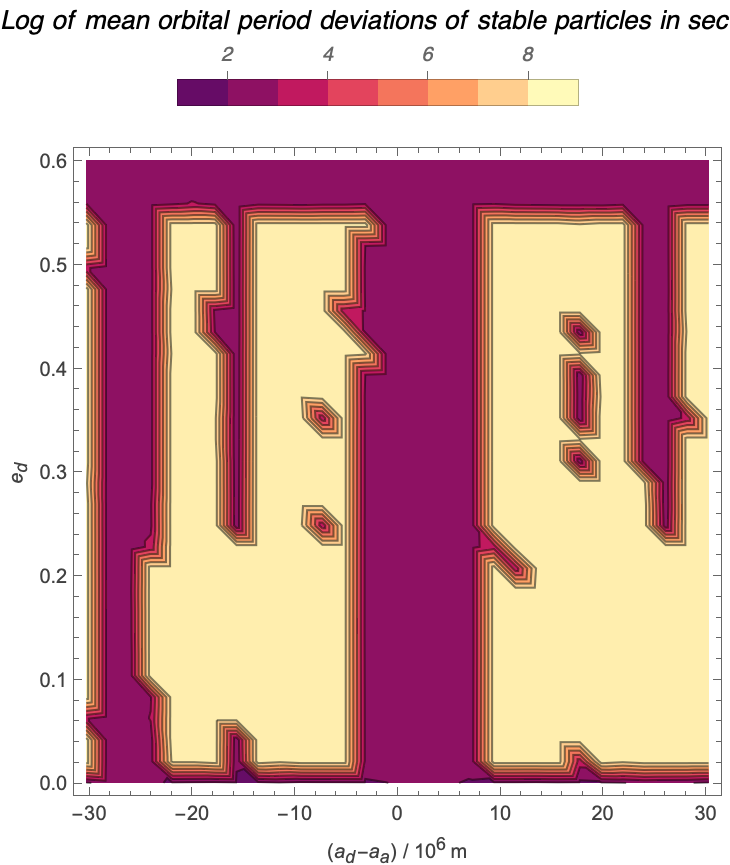}
\end{array}$
\caption{Simulations with initial conditions  extracted by an unstable periodic orbit of the $G^{II}$ family with  $\Delta\Omega=0^{\circ}$ and $\Delta i=85^{\circ}$ on the $(a_{\rm d} - a_{\rm a},e_{\rm d})$ plane. Presentation is the same as in Fig. \ref{sFI_we}.}
\label{sGII_aer}
\end{figure*}

Figure \ref{sFI_we} illustrates outcomes for $\Delta i=25^{\circ}$ and $\Delta\Omega=270^{\circ}$ on the $(\omega_{\rm d},e_{\rm d})$ plane. Then, in Fig. \ref{sFII_wel} we give the results for $\Delta\Omega=90^{\circ}$ and $\Delta i=10^{\circ}$, while in Fig. \ref{sFII_wer}, we give the results for $\Delta\Omega=90^{\circ}$ and $\Delta i=45^{\circ}$. Furthermore, in Figs. \ref{sFII_ael}-\ref{sFII_aer}, we project the same simulation outputs on the $(a_{\rm d} - a_{\rm a},e_{\rm d})$ plane. In Figs. \ref{sGII_ael}-\ref{sGII_aer}, we investigate values of $\Delta\Omega=0^{\circ}$ and $\Delta i=10^{\circ}$ as well as $\Delta\Omega=0^{\circ}$ and $\Delta i=85^{\circ}$ on the $(a_{\rm d} - a_{\rm a},e_{\rm d})$ plane, while the same output on the $(\omega_{\rm d},e_{\rm d})$ plane is provided in Figs. \ref{sGII_wel}-\ref{sGII_wer}.

\subsection{Collisions}\label{cols}

\begin{figure*}[t]
\centering
$\begin{array}{c}
\includegraphics[width=0.84\textwidth]{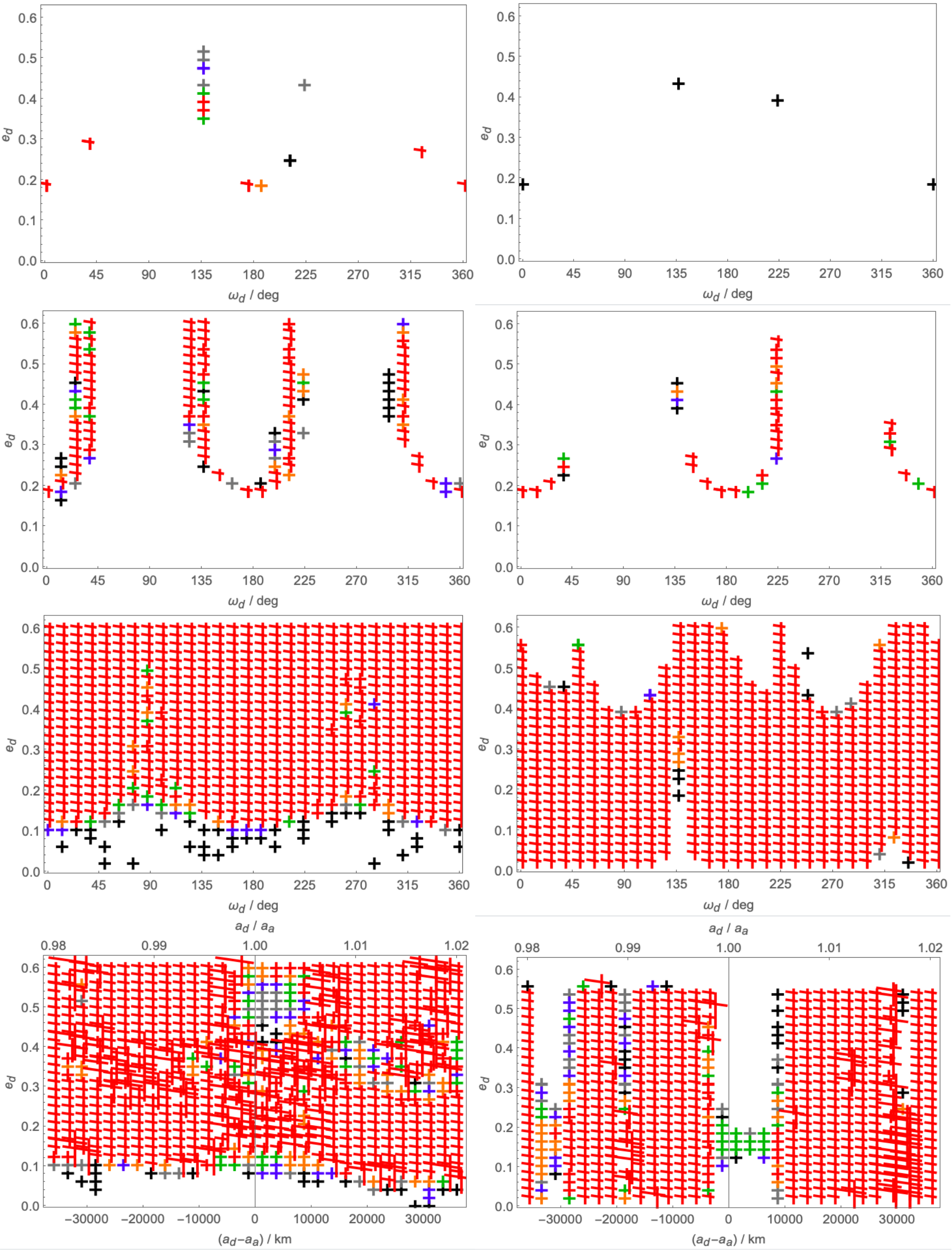}\\
\end{array}$
\caption{Collisions on the $(\omega_{\rm d},e_{\rm d})$ plane with Mercury (\textit{top}), Earth (\textit{centre-top}), and Neptune (\textit{centre-bottom}) for $\Delta i=10^{\circ}$ (\textit{left}) and $\Delta i=85^{\circ}$ (\textit{right}), and on the $(a_{\rm d},e_{\rm d})$ plane (\textit{bottom}) for $\Delta i=10^{\circ}$ (\textit{left}) and $\Delta i=85^{\circ}$ (\textit{right}) guided by periodic orbits of the $G^{II}$ family with $\Delta \Omega=0^{\circ}$. The symbol colours and shapes indicate how many of the 18 test particles for each initial condition collided with the planet: $(\{1,2,3,4,5,6,>6\}$ $= \{{\rm black, grey, blue, green, orange, red, red \,slanted}\})$.
}
\label{cMENweae}
\end{figure*}

All collisions occurred between the dust and the secondary. In Fig. \ref{cMENweae}, we indicate the incidence of direct collisions with the Mercury-, Earth-, and Neptune-like planets when the nodal difference is $\Delta\Omega=0^{\circ}$ on the $(\omega_{\rm d},e_{\rm d})$ and the $(a_{\rm d} - a_{\rm a},e_{\rm d})$ planes. A low ($\Delta i=10^{\circ}$; left panels) and very high ($\Delta i=85^{\circ}$, right panels) mutual inclination value was selected from the $G^{II}$ family.

\section{Discussion}\label{dis}

Figures \ref{sFI_we}--\ref{sGII_aer} demonstrate that over wide swathes of phase space, orbital period deviations due to planet-like objects (as opposed to asteroidal-type objects) are high enough to be observable. Photometric transit observations of WD planetary systems of the order of a few seconds have been realised from the ground \citep{gan16}. Hence, in principle, we ought to be able to discern variations that correspond to any symbol on these figures, except for the hollow downwards-pointing green triangles.

At the other extreme, the largest period deviations ($10^4$~s) often coincide with the highest incidence of collisions for different initial mean anomalies (Fig. \ref{cMENweae}). To explore this extreme even further, we ran additional simulations with secondaries that correspond to Jupiter-mass planets and brown dwarfs. The result was a nearly complete clear-out of the dust particles. Consequently, just the presence of transiting debris likely indicates that any hidden co-orbital exoplanet would be less massive than Jupiter.

The breakup of asteroids, and particularly planets, at the Roche sphere of the WD, can generate many massive fragments. Therefore, although we have focussed on the RTBP in this work, as well as in P3 and P5, we are aware that in some systems, four-body, five-body, or higher multiplicity body treatments could be more appropriate. \citep{vmtg16} investigated systems with four, six, and eight equal-mass massive bodies on coplanar orbits close to the Roche radius of a white dwarf, but with no dust. They ran $N$-body simulations and found that these massive bodies produce period deviations that are comparable to those given in Figs. \ref{sFI_we}-\ref{sGII_aer}.

\section{Conclusions}\label{concl}

In this work, we computed both known and new families of periodic orbits in the 1/1 MMR in the 2D-CRTBP and the 3D-CRTBP. New branches of highly eccentric asymmetric (TP) periodic orbits, called $A_2$ and $A_3$, were discovered in the 2D-CRTBP. The new families were generated by a  bifurcation point at $e_{\rm d}=0.982$, where the linear stability changes again along the family of HS orbits. The significance of the second new bifurcation point, $B_T^4$ found along the new $A_2$ family, is showcased in P5. We also reported the existence of two new families in the 3D-CRTBP that were generated by the circular family at 1/1 MMR. We applied these results to WD, secondary, and dust particle systems, extending the investigation of P3 to higher mass parameters corresponding to Ceres, Mercury, Earth, and Neptune.

We showcase the strong link between periodic orbits, DS maps, and simulations. Stable dust particles produced by the $N$-body code reside around stable periodic orbits and from a dynamical point of view, potentially forming observable debris discs or rings around WDs. Their orbital period deviations provide observational hints of hidden planets that are very close to WDs. 

More specifically, we studied three cases of nodal difference ($\Delta\Omega=0^{\circ},90^{\circ},\; {\rm and}\,270^{\circ}$) for distinct mutual inclination values ($\Delta i=10^{\circ}, 25^{\circ}, 45^{\circ}\,{\rm and}\, 85^{\circ}$). Most perturbers considered in this work lead to detectable orbital period changes in the orbits of the dust particles when they are fully outside (here at $a_{\rm a}=0.012~{\rm au})$ of the Roche radius; at least for low values of $e_{\rm d}$. Finally, we conclude that:\vspace{-0.2cm}

\begin{enumerate}
\item Planets that are co-orbital with dust produce easily detectable orbital period deviations over a decade.
\item As the planet mass increases, the planet clears away an increasing amount of dust at particular locations in the phase space.
\item In general, orbital period deviations are weakly dependent on inclination and highly dependent on the planet's mass.
\item For given values of the mass parameter and $\Delta\Omega$, the highly eccentric orbits of the dust were destabilised as the inclination of those orbits increased. 
\item Cases of high inclination (regardless of the nodal difference) yielded significantly greater orbital period deviations.
\item In the presence of WD debris, any hidden co-orbital exoplanet on a circular orbit is expected to be less massive than Jupiter.
\end{enumerate}\vspace{-0.2cm}
In the companion paper, P5, we take into account non-zero eccentricity values for Ceres, Mercury, Earth, and Neptune. We also explore how doing so would impact the results presented here.

\begin{acknowledgements} We thank the reviewer for constructive comments that enhanced the presentation of the results. This work started when KIA was supported by the University of Padua under Grant No. BIRD232319. Results have been produced using the Aristotle University of Thessaloniki HPC Infrastructure and Resources.
\end{acknowledgements}\vspace{-0.5cm}

\bibliographystyle{aa}
\bibliography{11}

\onecolumn
\begin{appendix}
\section{Supplementary material}\label{appsec}
\begin{table*}[h]
\centering
\caption{Coupled studies (periodic orbits and $N$-Body simulations) on WD pollution, where the WD is the primary body.}
\begin{tabular}[b]{cccccccc}
\toprule
Secondary
& Third body 
&MMR
& \begin{tabular}{@{}c@{}}Secondary on \\ Circular \\Orbit\end{tabular} 
& \begin{tabular}{@{}c@{}}Secondary on\\ Elliptic \\Orbit\end{tabular}  &\begin{tabular}{@{}c@{}}Mutual \\ Inclination\end{tabular} 
&Model  
&\begin{tabular}{@{}c@{}}Abbreviation \\ of study\end{tabular} \\
\cmidrule{1-8}
Jupiter & Asteroid & $2/1$ & $\checkmark$ &&&2D-CRTBP&P1 \\
\cmidrule{1-8}
Jupiter & Asteroid & $2/1$ 
&\begin{tabular}{@{}c@{}}$\checkmark$\\ \; \\ \; \end{tabular}  
&\begin{tabular}{@{}c@{}} \; \\$\checkmark$ \\$\checkmark$\end{tabular}
&\begin{tabular}{@{}c@{}}$\checkmark$\\ \; \\$\checkmark$\end{tabular}
&\begin{tabular}{@{}c@{}}3D-CRTBP \\2D-ERTBP \\ 3D-ERTBP\end{tabular}
&P2\\
\cmidrule{1-8}
$1/10^{\rm th}$ of Ceres & Dust particle & $1/1$ 
& \begin{tabular}{@{}c@{}}$\checkmark$ \\ $\checkmark$ \end{tabular}
&
&\begin{tabular}{@{}c@{}}\; \\ $\checkmark$ \end{tabular}
&\begin{tabular}{@{}c@{}}2D-CRTBP \\ 3D-CRTBP \end{tabular}
&P3 \\
\cmidrule{1-8}
\begin{tabular}{@{}c@{}}Ceres\\ Mercury \\Earth \\Neptune\end{tabular} & Dust particle & $1/1$ 
& \begin{tabular}{@{}c@{}}$\checkmark$ \\ $\checkmark$ \end{tabular}
&
&\begin{tabular}{@{}c@{}}\; \\ $\checkmark$ \end{tabular}
&\begin{tabular}{@{}c@{}}2D-CRTBP \\ 3D-CRTBP \end{tabular}
&\begin{tabular}{@{}c@{}}P4 \\ (This work) \end{tabular}  \\
\cmidrule{1-8}
\begin{tabular}{@{}c@{}}Ceres\\ Mercury \\Earth \\Neptune\end{tabular} & Dust particle & $1/1$ 
& 
& \begin{tabular}{@{}c@{}}$\checkmark$ \\ $\checkmark$ \end{tabular}
&\begin{tabular}{@{}c@{}}\; \\ $\checkmark$ \end{tabular}
&\begin{tabular}{@{}c@{}}2D-ERTBP \\ 3D-ERTBP \end{tabular}
&P5\\
\bottomrule
\end{tabular}\label{tab0}
\tablefoot{The checkmark-symbols are associated with the TBP configuration of the model between the secondary and the third body in each study.}
\tablebib{
(P1)~\citet{ave16}; (P2) \citet{ave19}; (P3) \citet{ave24}; (P4; This work);
(P5) \citet{ave26b}.}
\end{table*}
\FloatBarrier
\begin{figure*}[!h]
\centering
\sidecaption
\includegraphics[width=12cm]{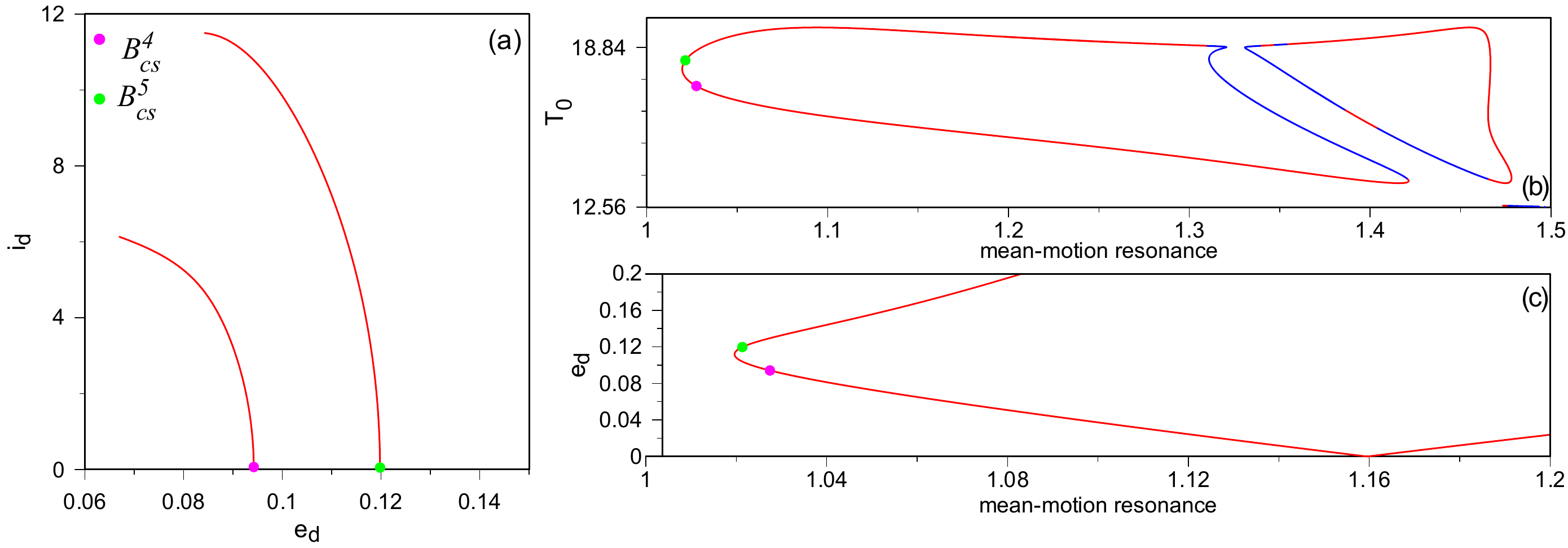}
\caption{(a) Families of 3D unstable periodic orbits, which bifurcate from the circular family (with circular periodic orbits; first species and first kind according to \citet{hen97} in the 2D-CRTBP. (b) Segment of the circular family below 3/2 MMR. $T_0$ stands for the period of the asteroid. (c) Magnification of (b) in relation to the eccentricity of the dust. $B_{cs}^4$ generates $xz$-symmetric periodic orbits in the 3D-CRTBP, while $B_{cs}^5$ generates $x$-symmetric ones.}
\label{circfam}
\end{figure*}
\FloatBarrier

\begin{table*}[h!]
\centering
\caption{Bifurcation points that exist in the CRTBP and justify the origin and continuation of all 1/1 resonant periodic orbits in the CRTBP presented within the Ceres-Neptune domain.}
\begin{tabular}[b]{lccccccccc}
\toprule
& $e_{\rm a}$ &$e_{\rm d}$&$\theta$  &$\varpi_{\rm a} (^{\circ})$     &$M_{\rm a} (^{\circ})$ &$\varpi_{\rm d} (^{\circ})$  &$M_{\rm d} (^{\circ})$& \begin{tabular}{@{}c@{}}Exists in \\ Problem/ \\ Plot(s)\end{tabular} & \begin{tabular}{@{}c@{}}Generates \\orbits in \\ Problem/Plot\end{tabular}\\
\cmidrule{1-10}
$B^1_{cs}$    &0      &0.697   &0     &0       &0   &180   &180 & \begin{tabular}{@{}c@{}}2D-CRTBP \\ Fig. \ref{2DC}a, \ref{2DC_all}\end{tabular} & \begin{tabular}{@{}c@{}}3D-CRTBP \\ Fig. \ref{3DC}a \end{tabular}  \\
\cmidrule{1-10}
$B^2_{cs}$    &0      &0.707   &180     &0       &0   &0   &180 & \begin{tabular}{@{}c@{}}2D-CRTBP \\ Fig. \ref{2DC}b, \ref{2DC_all}\end{tabular} & \begin{tabular}{@{}c@{}}3D-CRTBP \\ Fig. \ref{3DC}b\end{tabular}\\
\cmidrule{1-10}
$B^3_{cs}$    &0      &0.903   &180     &0       &0   &0   &180 & \begin{tabular}{@{}c@{}}2D-CRTBP \\ Fig. \ref{2DC}c, \ref{2DC_all}\end{tabular} & \begin{tabular}{@{}c@{}}3D-CRTBP \\ Fig. \ref{3DC}c\end{tabular}\\
\cmidrule{1-10}
$B^4_{cs}$    &0      &0.094   &0     &0       &0   &0   &0 & \begin{tabular}{@{}c@{}}2D-CRTBP \\ Fig. \ref{circfam}b,c\end{tabular} & \begin{tabular}{@{}c@{}}3D-CRTBP \\ Fig. \ref{circfam}a\end{tabular}\\
\cmidrule{1-10}
$B^5_{cs}$    &0      &0.119   &0     &0       &0   &0   &0 & \begin{tabular}{@{}c@{}}2D-CRTBP \\ Fig. \ref{circfam}b,c\end{tabular} & \begin{tabular}{@{}c@{}}3D-CRTBP \\ Fig. \ref{circfam}a\end{tabular}\\
\cmidrule{1-10}
$B^1_{T}$    &0      &0.835   &0     &0       &0   &180   &180 & \begin{tabular}{@{}c@{}}2D-CRTBP \\ Fig. \ref{2DC}a, \ref{2DC_all}\end{tabular} & \begin{tabular}{@{}c@{}}2D-ERTBP \end{tabular}\\
\cmidrule{1-10}
$B^2_{T}$    &0      &0.977   &180     &0       &0   &0   &180 & \begin{tabular}{@{}c@{}}2D-CRTBP \\ Fig. \ref{2DC}b, \ref{2DC_all}\end{tabular} & \begin{tabular}{@{}c@{}}2D-ERTBP \end{tabular}\\
\cmidrule{1-10}
$B^3_{T}$    &0      &0.869   &146.74     &180       &180   &133.43   &79.83 & \begin{tabular}{@{}c@{}}2D-CRTBP \\ Fig. \ref{2DC}c, \ref{2DC_all}\end{tabular} & \begin{tabular}{@{}c@{}}2D-ERTBP \end{tabular}\\
\cmidrule{1-10}
$B^4_{T}$    &0      &0.964   &167.19     &180       &180   &314.66   &212.53 & \begin{tabular}{@{}c@{}}2D-CRTBP \\ Fig. \ref{2DC_all}\end{tabular} & \begin{tabular}{@{}c@{}}2D-ERTBP \end{tabular}\\
\cmidrule{1-10}
$B^5_{T}$    &0      &\begin{tabular}{@{}c@{}}0.815 \\ (when $i_{\rm d}=22.5^{\circ}$)\end{tabular}   &0      &\begin{tabular}{@{}c@{}}0 \\ (with $\Omega_{\rm a}=0^{\circ}$\\ and $\omega_{\rm a}=0^{\circ}$)\end{tabular}     &0       &\begin{tabular}{@{}c@{}}180 \\ (with $\Omega_{\rm d}=270^{\circ}$\\ and $\omega_{\rm d}=270^{\circ}$)\end{tabular}
   &180 & \begin{tabular}{@{}c@{}}3D-CRTBP \\ Fig. \ref{3DC}a\end{tabular} & \begin{tabular}{@{}c@{}}3D-ERTBP \end{tabular}  \\
\bottomrule
\end{tabular}\label{tab1}
\tablefoot{As the order of $\mu$ varies from Ceres to Neptune, no significant change was reflected on $e_{\rm d}$ (of the order of $10^{-4}$ for all bifurcation points). The plots regarding the 2D-ERTBP and 3D-ERTBP are included in the respective Table of P5.}
\end{table*}
\FloatBarrier

\begin{figure*}[!h]\centering
$\begin{array}{c}
\includegraphics[width=0.99\textwidth]{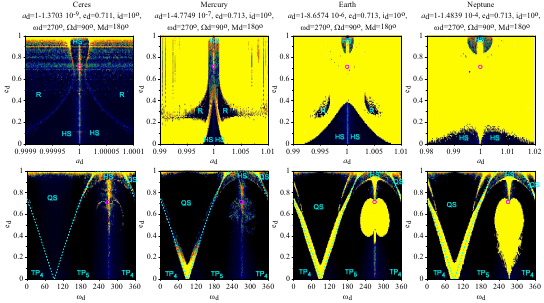}\\\includegraphics[width=0.99\textwidth]{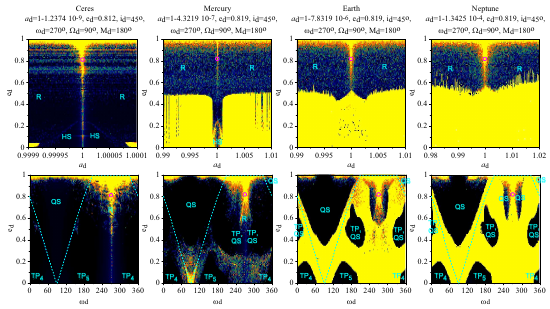}\vspace{-0.2cm}\\
\hspace{0.3cm}\includegraphics[width=0.18\textwidth]{bar.pdf}\end{array}$
\caption{Pairs of DS maps, presented as in Fig. \ref{mFI}, for $\Delta i=10^{\circ}$  (\textit{top}) and $\Delta i=45^{\circ}$ (\textit{bottom}) guided by an unstable periodic orbit (magenta-coloured circle) of the $F^{II}$ family. The resonant angle can be computed via $\omega_{\rm d}$, as $\theta=-\omega_{\rm d}-270$.}\vspace{-0.3cm}
\label{mFII}
\end{figure*}

\begin{figure*}[!h]\centering
$\begin{array}{c}
\includegraphics[width=0.99\textwidth]{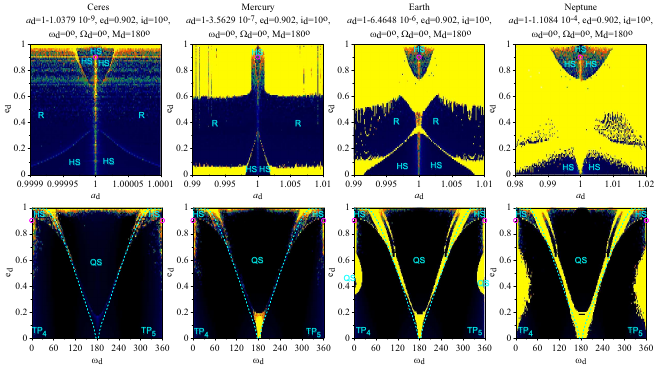}\\\includegraphics[width=0.99\textwidth]{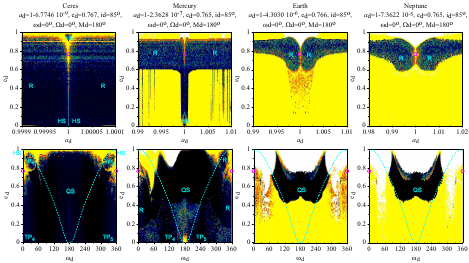}\vspace{-0.2cm}\\
\hspace{0.3cm}\includegraphics[width=0.18\textwidth]{bar.pdf}\end{array}$
\caption{Pairs of DS maps, presented as in Fig. \ref{mFI}, for $\Delta i=10^{\circ}$  (\textit{top}) and $\Delta i=85^{\circ}$ (\textit{bottom}) guided by an unstable periodic orbit  (magenta-coloured circle) of the $G^{II}$ family.  The resonant angle can be computed via $\omega_{\rm d}$, as $\theta=-\omega_{\rm d}-180$.}\vspace{-0.3cm}
\label{mGII}
\end{figure*}

\begin{figure*}\centering
$\begin{array}{c}
\includegraphics[width=0.99\textwidth]{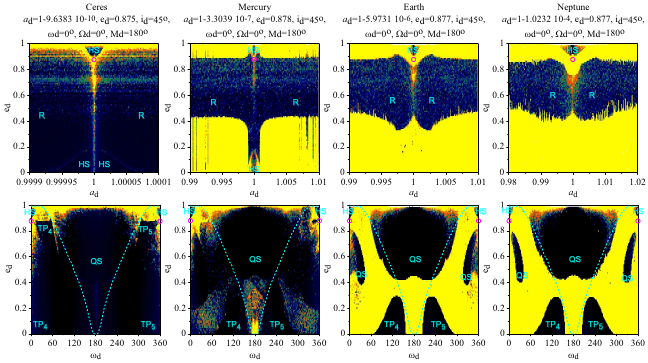}\vspace{-0.1cm}\\
\includegraphics[width=0.18\textwidth]{bar.pdf}\end{array}$\vspace{-0.5cm}
\caption{Same as in Fig. \ref{mGII} but for $\Delta i=45^{\circ}$.}\vspace{-0.1cm}
\label{2DC_S10}
\end{figure*}

\begin{figure*}\centering
$\begin{array}{c}
\includegraphics[width=0.8\textwidth]{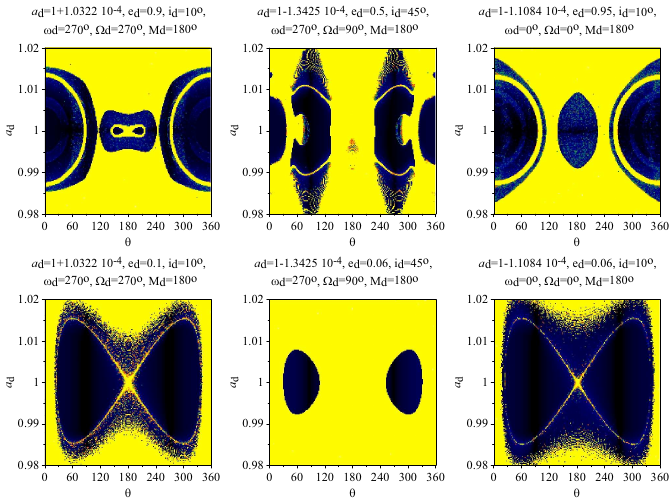}\\
\includegraphics[width=0.18\textwidth]{bar.pdf}\end{array}$\vspace{-0.2cm}
\caption{DS maps on ($\theta,a_{\rm d}$) plane that justify the classification (QS, HS, and TP$_{4,5}$) used in Fig. \ref{mFI} and Figs. \ref{mFII}--\ref{2DC_S10}. Initial conditions are reported above each map spanning over a 200$\times$200 grid, and although they correspond to a Neptune-like secondary, they are typical librations or rotations encountered for the rest $\mu$ values considered in this study.}\vspace{-0.2cm}
\label{thetaa}
\end{figure*}
\FloatBarrier

\begin{figure*}[!h]\centering
\sidecaption
$\begin{array}{cc}
\includegraphics[width=6cm]{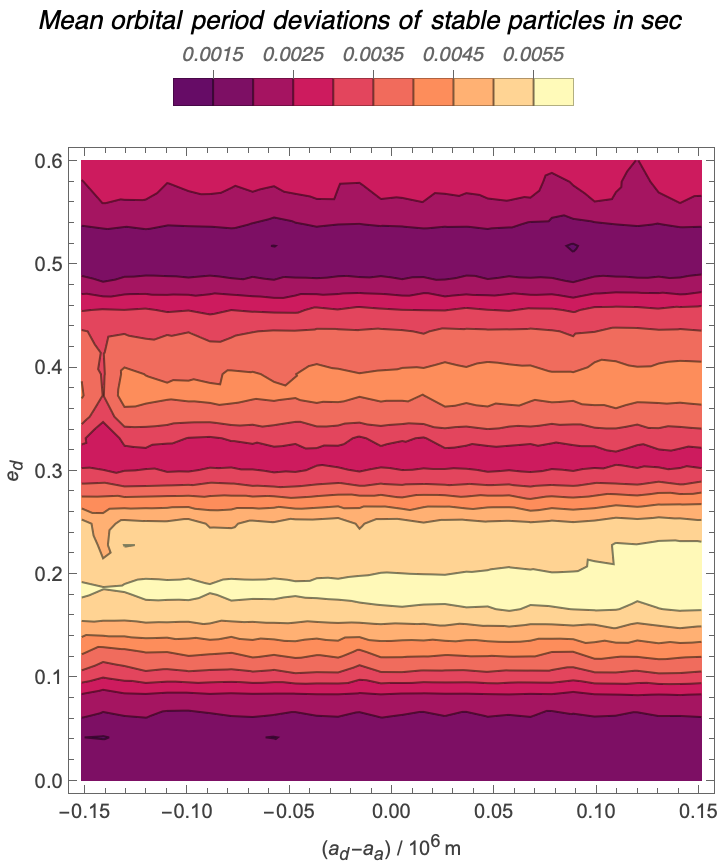}& \includegraphics[width=6cm]{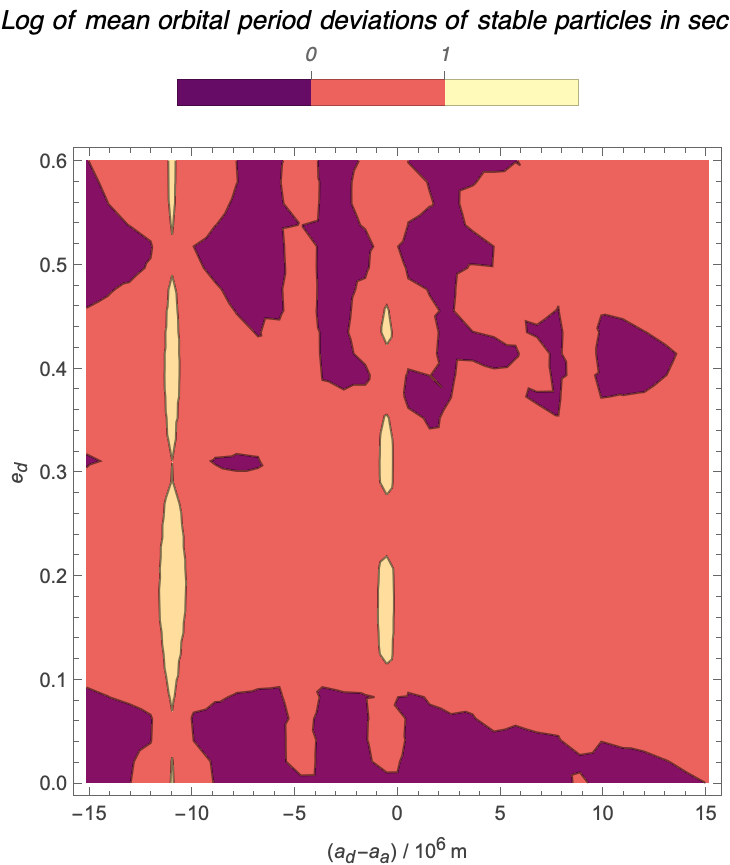}\\\includegraphics[width=6cm]{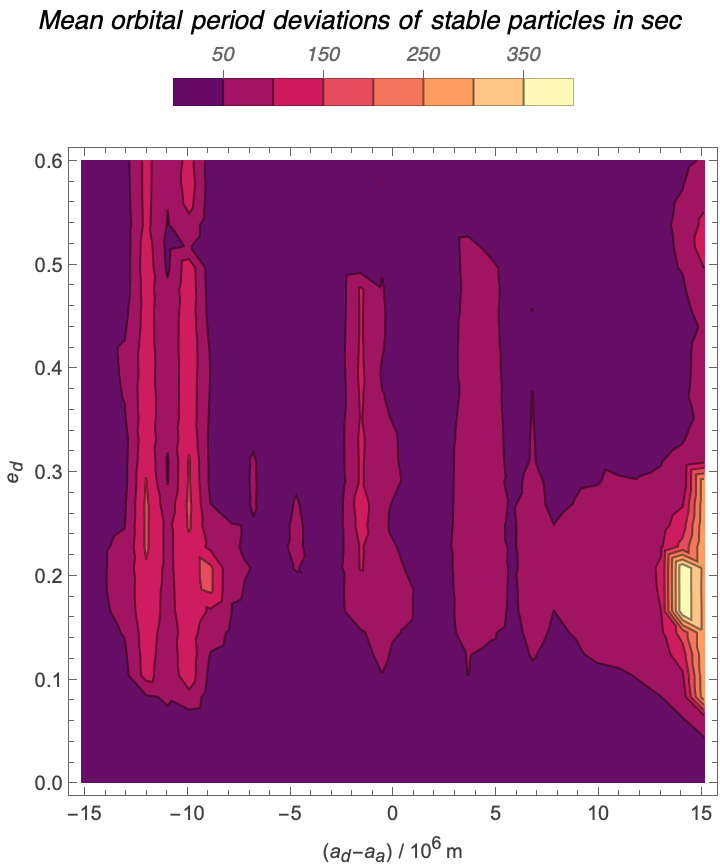}& \includegraphics[width=6cm]{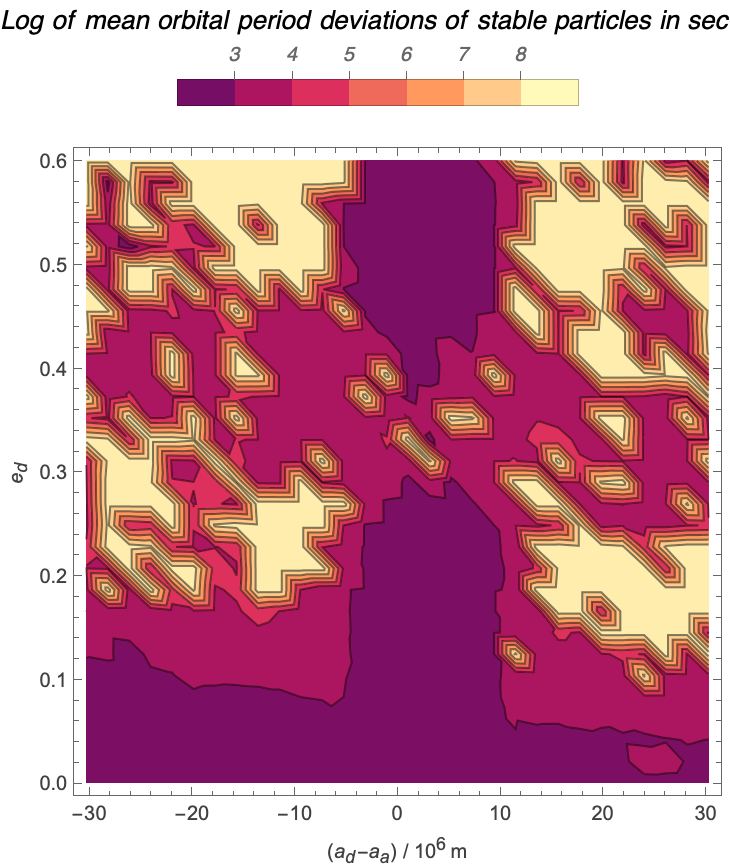}
\end{array}$\vspace{-0.5cm}
\caption{Simulations over a 10 yr time span on the $(a_{\rm d} - a_{\rm a},e_{\rm d})$ plane. The initial conditions were extracted by an unstable periodic orbit of the $F^{II}$ family with  $\Delta\Omega=90^{\circ}$ and $\Delta i=10^{\circ}$. Presentation is the same as in Fig. \ref{sFI_we}.}\vspace{-0.2cm}
\label{sFII_ael}
\end{figure*}
\FloatBarrier

\begin{figure*}[!h]\centering
\sidecaption
$\begin{array}{cc}
\includegraphics[width=6cm]{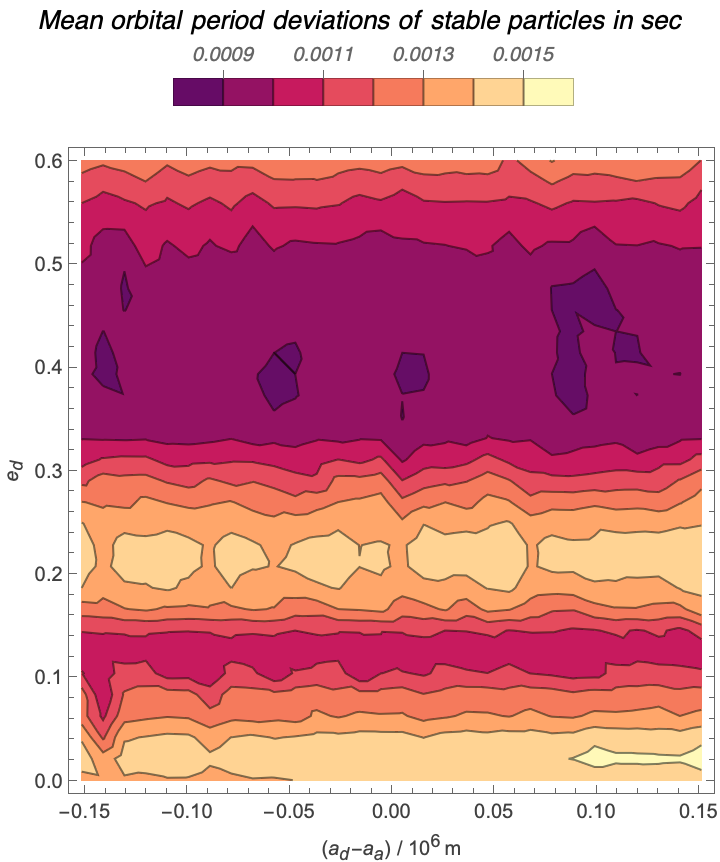}& \includegraphics[width=6cm]{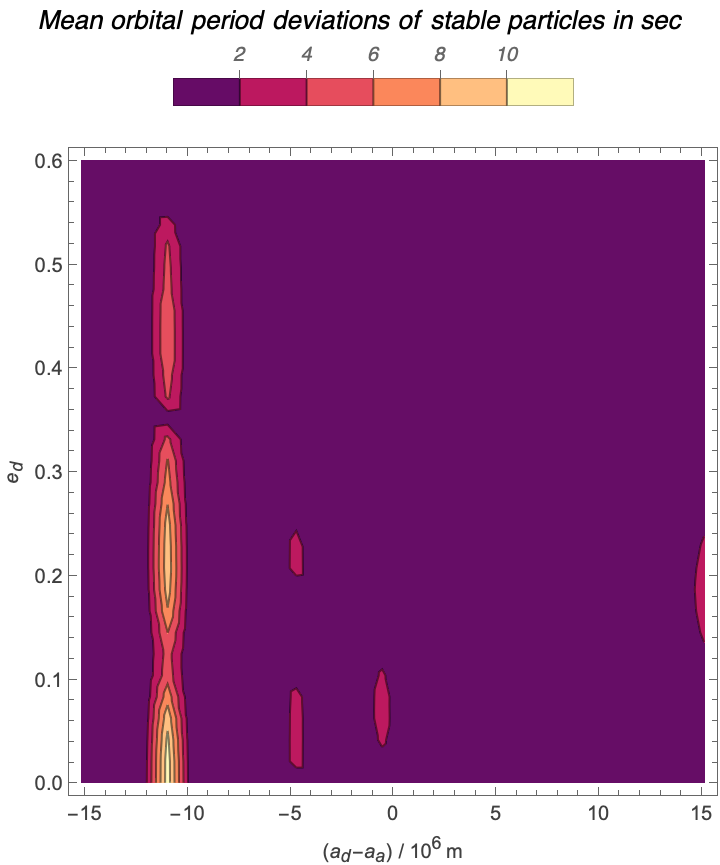}\\\includegraphics[width=6cm]{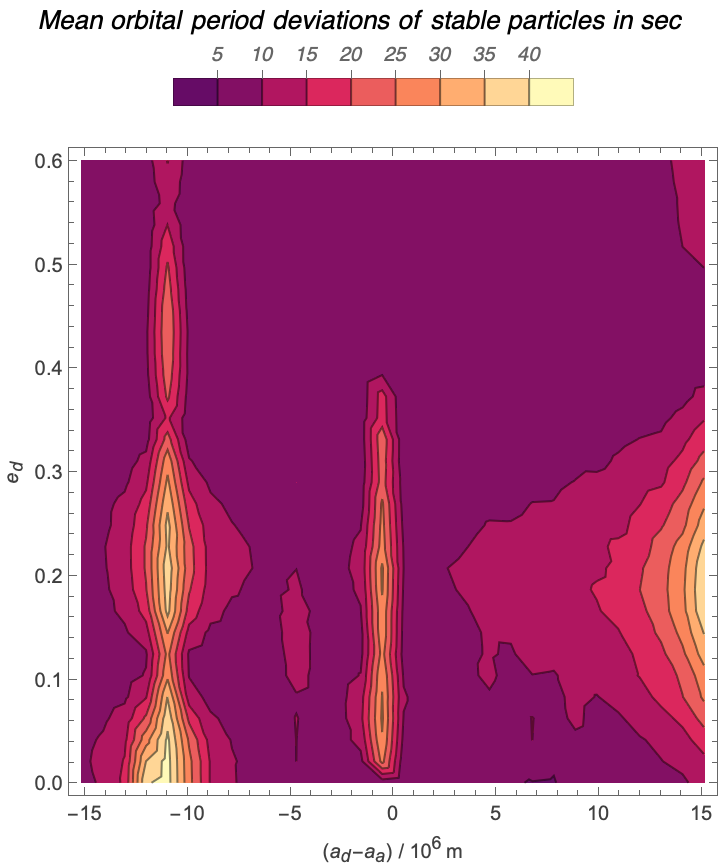}& \includegraphics[width=6cm]{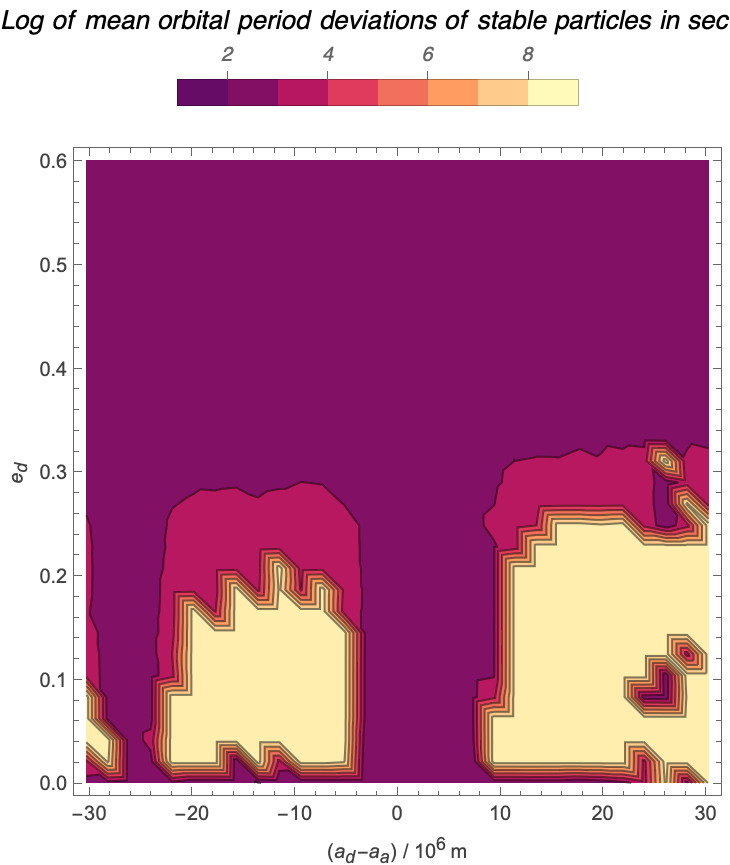}
\end{array}$
\caption{Simulations over a 10 yr timespan on the $(a_{\rm d} - a_{\rm a},e_{\rm d})$ plane. The initial conditions were extracted by an unstable periodic orbit of the $F^{II}$ family with  $\Delta\Omega=90^{\circ}$ and $\Delta i=45^{\circ}$ (\textit{right}). Presentation is the same as in Fig. \ref{sFI_we}.}\vspace{-0.5cm}
\label{sFII_aer}
\end{figure*}
\FloatBarrier

\begin{figure*}[!h]\centering
\sidecaption
$\begin{array}{cc}
\includegraphics[width=6cm]{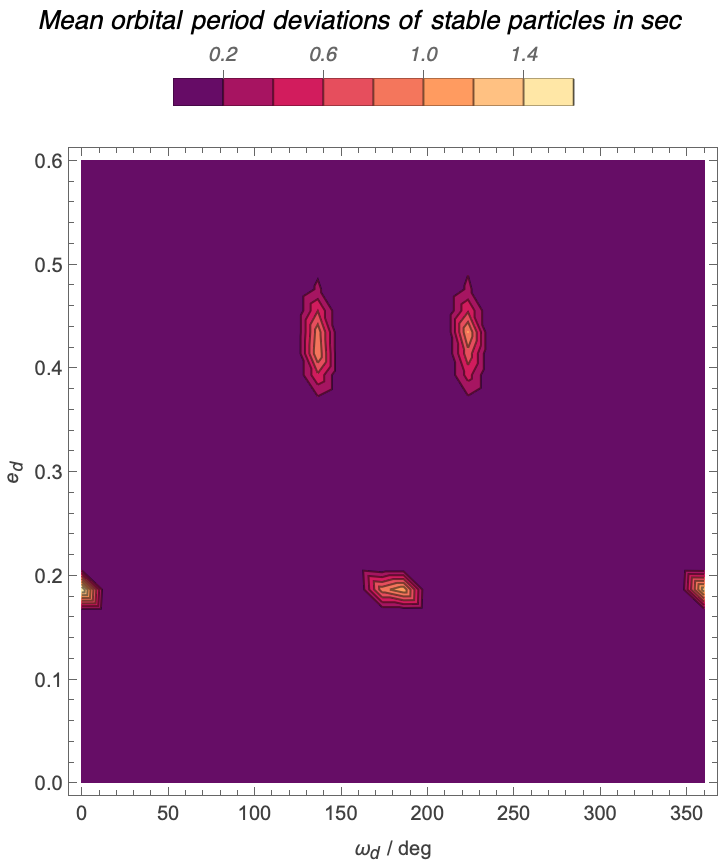}& \includegraphics[width=6cm]{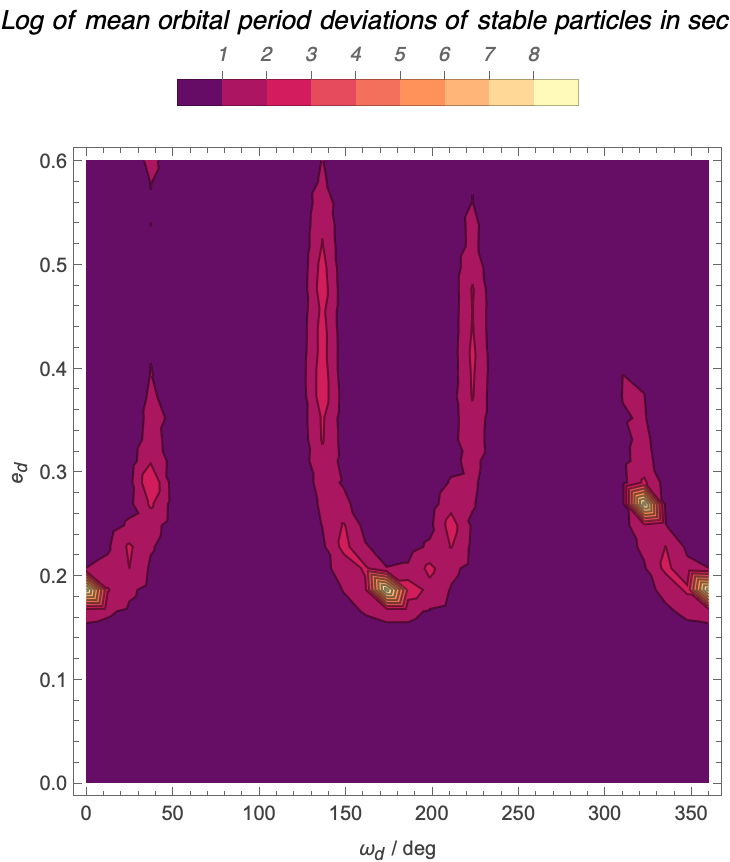}\\\includegraphics[width=6cm]{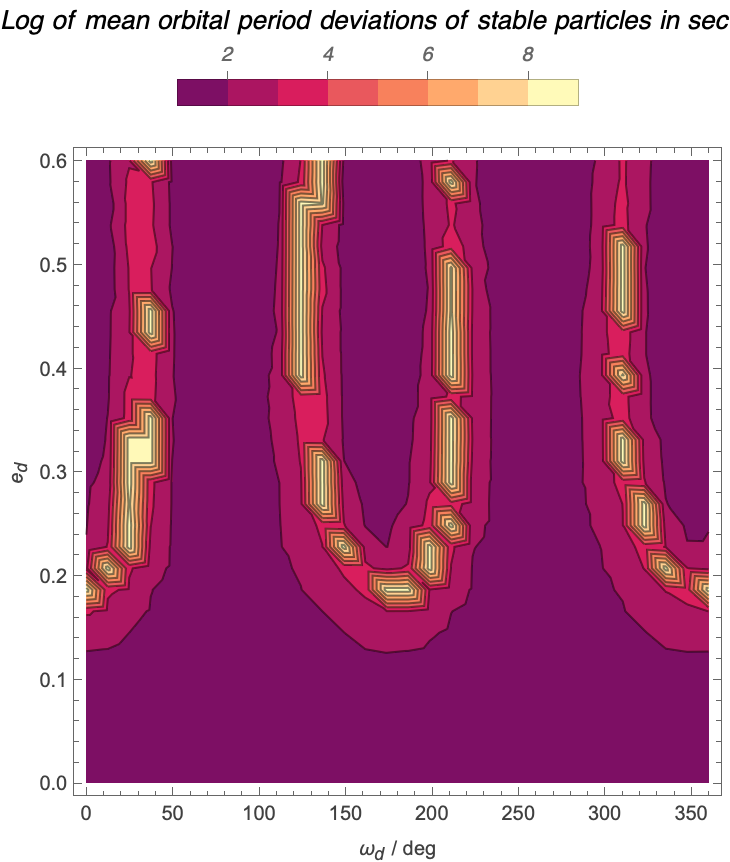}& \includegraphics[width=6cm]{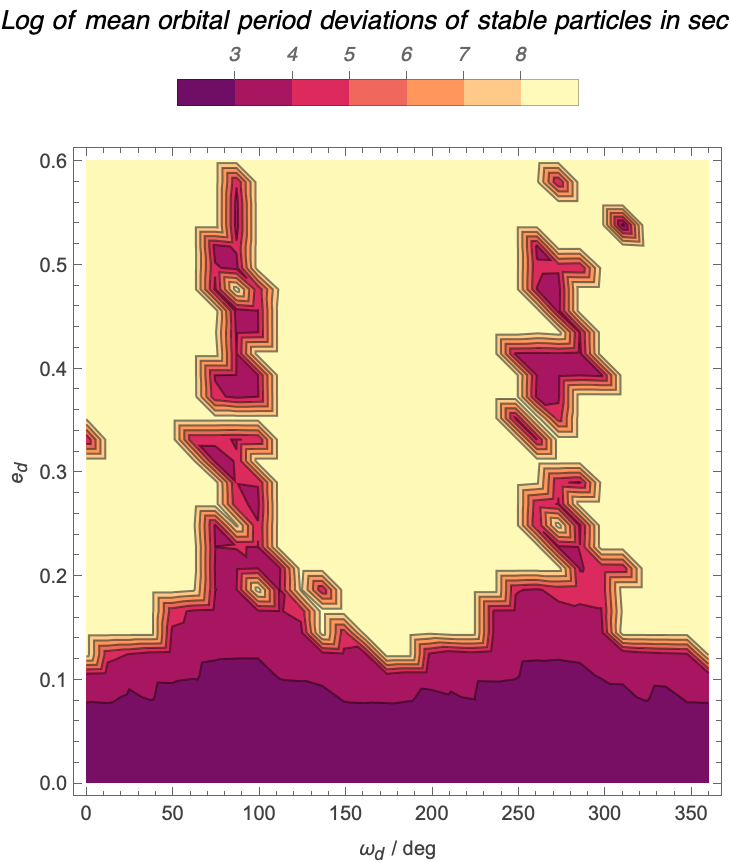}
\end{array}$
\caption{Simulations over a 10 yr time span on the $(\omega_d,e_d)$ plane. The initial conditions were extracted by an unstable periodic orbit of the $G^{II}$ family with $\Delta\Omega=0^{\circ}$ and $\Delta i=10^{\circ}$. Presentation is the same as in Fig. \ref{sFI_we}.}
\label{sGII_wel}
\end{figure*}
\FloatBarrier

\begin{figure*}[!h]\centering
\sidecaption
$\begin{array}{cc}
\includegraphics[width=6cm]{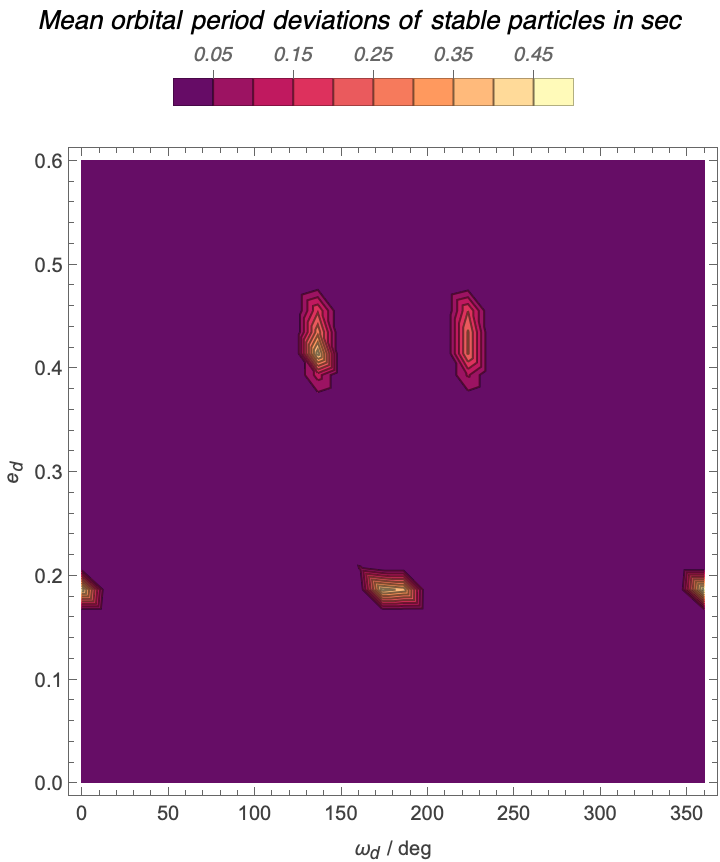}& \includegraphics[width=6cm]{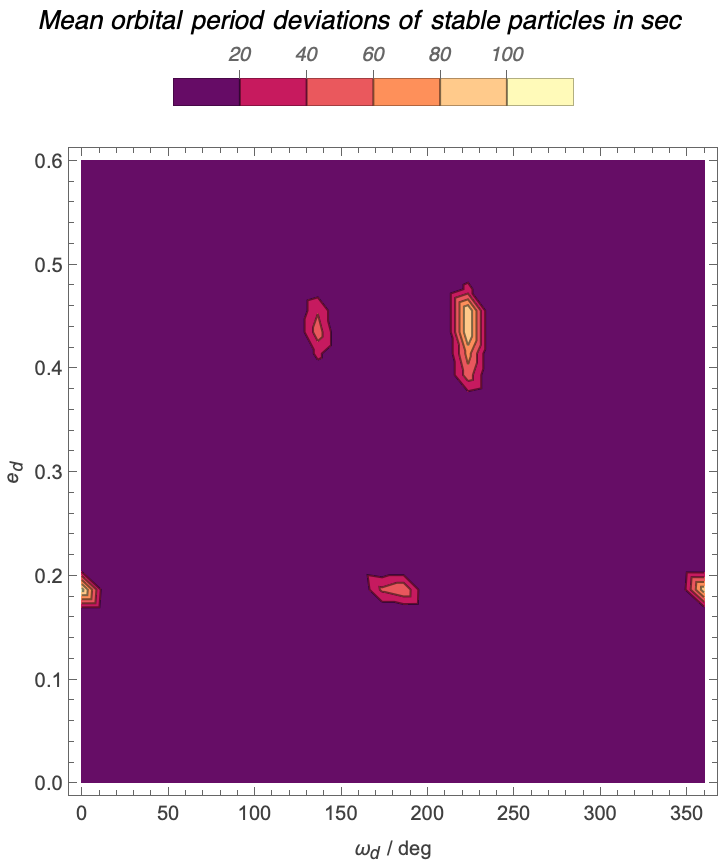}\\\includegraphics[width=6cm]{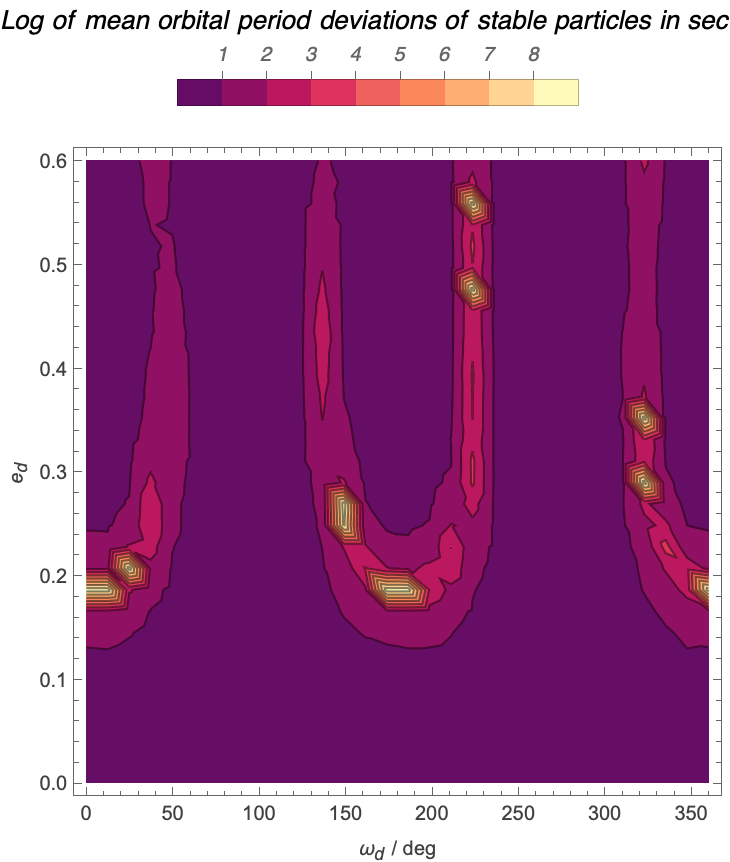}& \includegraphics[width=6cm]{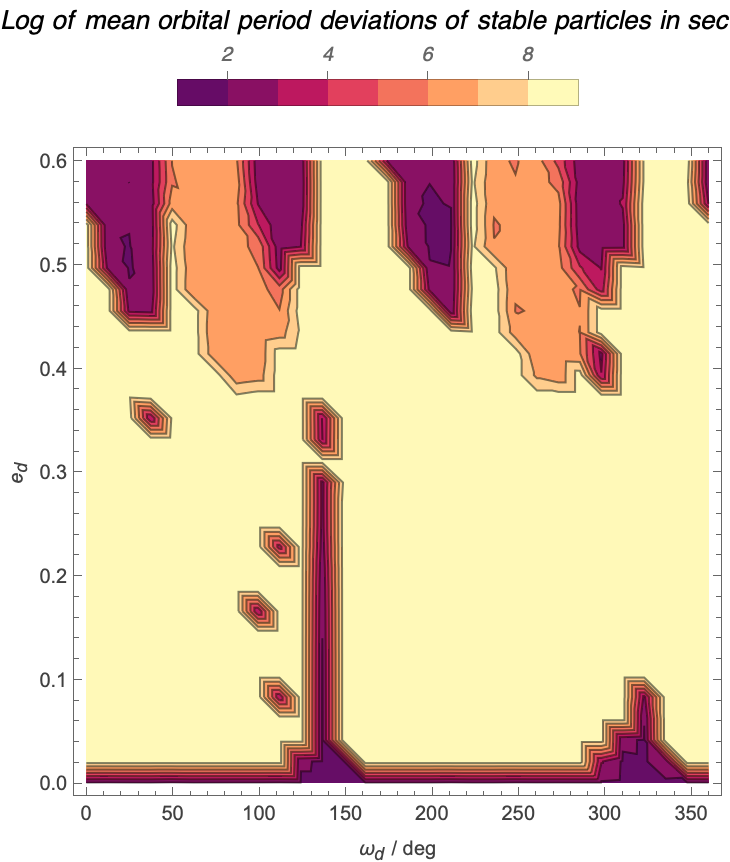}%
\end{array}$
\caption{Simulations over a 10 yr time span on the $(\omega_d,e_d)$ plane. The initial conditions were extracted by an unstable periodic orbit of the $G^{II}$ family with $\Delta\Omega=0^{\circ}$ and $\Delta i=85^{\circ}$. Presentation is the same as in Fig. \ref{sFI_we}.}
\label{sGII_wer}
\end{figure*}
\FloatBarrier

\end{appendix}

\end{document}